\documentclass[preprint,12pt]{elsarticle}

\usepackage{amssymb}
\usepackage{latexsym}
\usepackage{graphicx}
\usepackage{color}
\usepackage{natbib}
\usepackage{amssymb,amsfonts}
\usepackage{amsmath,accents}
\usepackage{textgreek}
\usepackage{soul,xcolor}
\journal{Additive Manufacturing}

\begin{document}

\begin{frontmatter}

\title{Resolving thermal gradients and solidification velocities during laser melting of a refractory alloy}

\author[inst1]{Hyunggon Park$\dagger$}

\affiliation[inst1]{organization={Department of Mechanical Engineering, University of California, Santa Barbara},%Department and Organization
            %addressline={Address One}, 
            %city={Santa Barbara},
            postcode={93106}, 
            state={CA},
            country={USA}}

\author[inst2]{Kaitlyn M. Mullin$\dagger$}
\author[inst1]{Vijay Kumar$\dagger$}
\author[inst2]{Olivia A. Wander}
%\author[inst2]{Raphaële J. Clément}
\author[inst2]{Tresa M. Pollock*}
\author[inst1]{Yangying Zhu*}

\affiliation[inst2]{organization={
Materials Department, University of California, Santa Barbara},%Department and Organization
            %addressline={Address Two}, 
            %city={City Two},
            postcode={93106}, 
            state={CA},
            country={USA}}

\begin{abstract}
 Metal additive manufacturing (AM) processes, such as laser powder bed fusion (L-PBF), can yield high-value parts with unique geometries and features, substantially reducing costs and enhancing performance. However, the material properties from L-PBF processes are highly sensitive to the laser processing conditions and the resulting dynamic temperature fields around the melt pool. In this study, we develop a methodology to measure thermal gradients, cooling rates, and solidification velocities during solidification of refractory alloy C103 using in situ high-speed infrared (IR) imaging with a high frame rate of approximately 15,000 frames per second (fps). Radiation intensity maps are converted to temperature maps by integrating thermal radiation over the wavelength range of the camera detector while also considering signal attenuation caused by optical parts.  Using a simple method that assigns the liquidus temperature to the melt pool boundary identified ex situ, a scaling relationship between temperature and the IR signal was obtained. The spatial temperature gradients ($dT/dx$), heating/cooling rates ($dT/dt$), and solidification velocities ($R$) are resolved with sufficient temporal resolution under various laser processing conditions, and the resulting microstructures are analyzed, revealing epitaxial growth and nucleated grain growth. Thermal data shows that a decreasing temperature gradient and increasing solidification velocity from the edge to the center of the melt pool can induce a transition from epitaxial to equiaxed grain morphology, consistent with the previously reported columnar to equiaxed transition (CET) trend. The methodology presented can reduce the uncertainty and variability in AM and guide microstructure control during AM of metallic alloys.

\end{abstract}

\begin{graphicalabstract}
\end{graphicalabstract}

\begin{highlights}

%\item We present a methodology to convert high-speed IR images of the melt pools during laser melting of a Nb-base alloy to temperature maps by integrating radiation over the wavelength range of the IR detector and by assigning the liquidus temperature to the melt pool boundary without sophisticated X-ray imaging.

\item We present a methodology to convert high-speed IR images during the laser melting of an Nb-base alloy to temperature maps.

% \item We present a method to resolve temperature gradients, cooling rates, melt pool dimensions, and solidification velocities with high temporal and spatial resolution.

\item We present a method to resolve thermal parameters with high temporal and spatial resolution.

%\item We observe temperature gradients in the range of 1.3--2\,$\times10^7$\,K/m and cooling rates up to 8\,$\times10^6$\,K/s along the melt pool boundary for C103 refractory alloy under the conditions tested. 

%\item During solidification, our thermal data show that a decreasing thermal gradient and increasing solidification velocity from the edge to the center of the melt pool can result in a transition from epitaxial to equiaxed grain morphology in a refractory alloy.

\item Our thermal data validate the transition from epitaxial to equiaxed grain morphology in a refractory alloy.

\end{highlights}


\end{frontmatter}



\section{Indroduction}
\label{sec:intro}

Laser powder bed fusion (L-PBF), also known as selective laser melting (SLM), is a widely utilized metal additive manufacturing (AM) technique that employs a laser beam to melt fine powders onto a substrate surface, enabling layer-wise and location-specific deposition \cite{kotadia2021review}. This method enables the production of intricate geometries and unconventional materials unattainable through conventional manufacturing, with site-specific control of material properties through tailored printing parameters \cite{raghavan2017localized,dehoff2015site}. The density and properties of the printed material are highly sensitive to the solidification conditions, including the dynamic temperature field. During L-PBF, the high-energy focused laser beam causes rapid melting and high thermal gradients. The resulting heat and fluid flow affect the melt pool dynamics and the cooling rate, which directly control grain growth and resulting microstructures \cite{david1989correlation}. Unfavorable thermal conditions can lead to high crack susceptibility \cite{collins2016microstructural,debroy2018additive}, lack of fusion \cite{das2003physical,echeta2020review}, distortion with surface roughness \cite{collins2016microstructural,echeta2020review}, hot tearing \cite{kimura2017effect}, balling \cite{attar2014manufacture,scime2019melt}, and keyholing \cite{cunningham2019keyhole,scime2019melt,zhao2020critical,ren2023machine}, which can all impact the final density and mechanical properties of the printed component. Previous efforts to optimize thermal processes in L-PBF have mostly relied on trial-and-error prints with varying laser parameters such as power and scan rate \cite{oliveira2020processing,lewandowski2016metal,hanzl2015influence}. However, only tuning laser parameters can still yield dramatically different as-printed densities, attributed to machine-to-machine variability. Although computational simulations are also used to estimate thermal gradients and solidification velocities during laser melting, these methods are often computationally expensive and can be inaccurate due to the complex multiphysics and multiscale phenomena involved in SLM as well as the lack of thermophysical material properties \cite{markl2016multiscale}. As such, microstructure control strategies have relied on variation of process parameters, coupled with computational simulations to estimate the thermal gradients and solidification velocities \cite{shi2020microstructural,kirka2017strategy}. Direct measurements of the temperature gradient, cooling rate, and solidification velocity during L-PBF are essential for repeatable microstructure control and defect mitigation \cite{raplee2017thermographic,gould2021situ,wang2022situ}. A framework for measuring these values will accelerate development of strategies for microstructure control in new materials during L-PBF.

Thermographic measurements in L-PBF offer valuable insights into the solidification conditions during the manufacturing process. These measurements aid in identifying defect signatures, such as keyholing \cite{ren2023machine}, microstructure prediction \cite{raplee2017thermographic}, and even recognizing dynamic phenomena such as vapor plume dynamics, spatter formation, and thermal history \cite{heigel2017effect,heigel2018measurement,gould2021situ,wang2022situ}. The complexity of thermal-fluid models also highlights the critical need for precise temperature measurements within and around the melt pool for model validation \cite{cheng2019computational,debroy2018additive}. For this, various thermographic methods have been utilized, including pyrometers \cite{tapia2014review,everton2016review}, infrared (IR) imaging \cite{heigel2018measurement,heigel2018measurement,heigel2020situ,gould2021situ,wang2022situ}, and two-color method with a color camera \cite{myers2023high,myers2023two}. While pyrometry usually provides a spatially averaged temperature, the two-color method is accompanied by a relatively high uncertainty ($\pm$\,100\,K). IR imaging has the potential to provide accurate temperature maps, but, it requires calibration of emissivity around the melting temperature, which is particularly challenging for high-temperature materials like refractory alloys 
\cite{raplee2017thermographic,wang2022situ}. Thus, IR techniques have mainly been used to either qualitatively capture hot spots, melt pool shapes and defects, or estimate temperature using emissivity values calibrated at a lower temperature.  More recently, high-speed X-ray imaging has been used synchronously with IR imaging to assign the liquidus temperature to the melt pool boundary identified by X-ray \cite{moylan2014infrared,cunningham2019keyhole,zhao2020critical,wang2022situ,ren2023machine}. The emissivity thus characterized was assumed constant around the melt pool to obtain temperature maps based on the Stefan-Boltzmann Law. Despite the advancement, such methods require sophisticated instrumentation in addition to approximating the temperature field because IR cameras do not capture radiation in the full wavelength spectrum as applied in the Stefan-Boltzmann Law. Consequently, there is still a need for a simple and accurate temperature measurement methodology. 

Additionally, many previous works have employed insufficient measurement frame rates, typically on the order of tens to thousands of frames per second, which are not fast enough to resolve cooling rates exceeding $10^6-10^7$\,K/s \cite{hooper2018melt,zagade2021analytical,wang2022situ}. The limited frame rates can also hinder the accurate measurement of melt pool width and increase measurement uncertainty \cite{heigel2018measurement,heigel2020situ,heigel2020situ2,raplee2017thermographic}, which underscore the necessity of high-frame-rate measurements. Thus, high-speed IR imaging ($>$10,000\,fps) is crucial for accurately resolving fast-moving thermal gradients and solidification processes, ensuring precise characterization of the thermal history \cite{gould2021situ,wang2022situ}.

High-temperature alloys, such as refractory alloys (alloyed with Cr, Hf, Mo, Nb, Re, Ru, Ta, Ti, V, W, and/or Zr), have benefited from recent advances in AM. AM can offer additional microstructural control and avoid challenges with conventional processing methods, such as cracking during themo-mechanical working. The Nb-base alloy, C103 (Nb-10Hf-1Ti, wt\,$\%$), has been the refractory alloy of choice for high temperature applications for a number of years due to its high fabricability \cite{philips2020new}. This ease in fabrication has also translated to AM, with numerous reports of fully dense printed C103 components with no cracking \cite{awasthi2022mechanical,colon2024parameter,mullin2024rapid,mireles2020additive,philips2024electron,miklas2022additive}. 
Determining processing-microstructure-property relationships is critical for advancing AM and enabling microstructure control, but remains in the early stages for C103.
For example, measuring mechanical properties and microstructural characteristics of the printed part \cite{awasthi2022mechanical,philips2024electron},  simulating the solidification behavior \cite{miklas2022additive}, and predicting the cracking susceptibility \cite{mullin2024rapid} have been investigated. However, microstructure predictions from simulations have not aligned with the experimental results \cite{mullin2024rapid} and changes in mechanical properties
are evident across printing conditions \cite{philips2024electron}. Understanding the thermal behavior during solidification is crucial for achieving the desired microstructure and properties in C103 and metallic alloys produced via AM.

\begin{figure*}
\centering
\includegraphics[width=.95\textwidth]{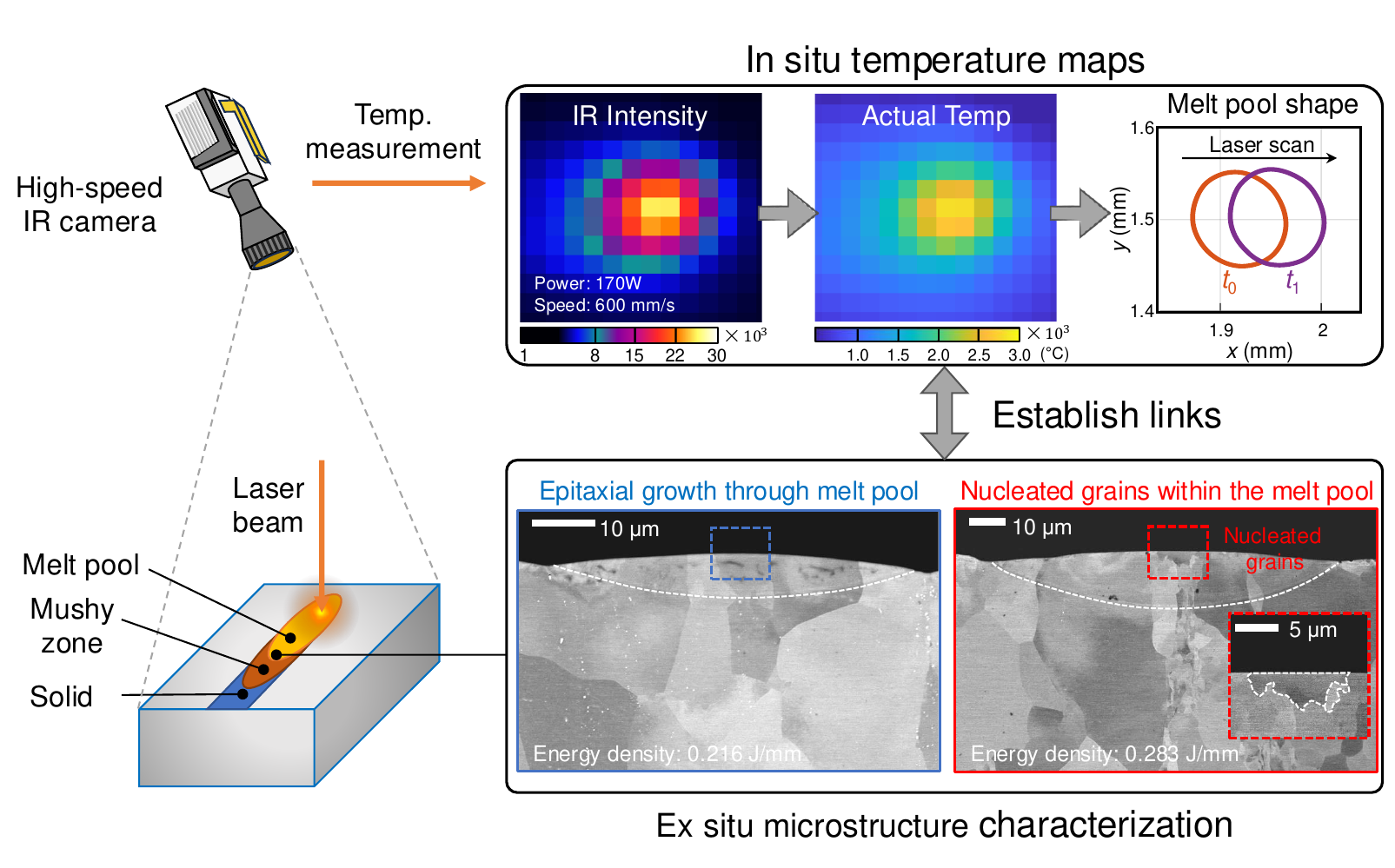}
\caption{Overview of the experimental procedure. A high-speed IR camera records radiation intensity maps which are then converted to the dynamic temperature maps during laser melting of C103. The temperature maps offer rich spatially and temporally resolved information such as the melt pool shape, the temperature gradient, the heating/cooling rate, and the solidification velocity. The microstructures of the melt pools are analyzed using SEM, indicating different grain growth dynamics under various laser processing parameters. Processing-microstructure relationships can be established by comparing thermal characteristics with the resulting microstructures.}
\label{fig:Overview}
\end{figure*}

Here, we report a methodology of converting in situ IR imaging to temperature maps near the melt pool of refractory alloy C103 without the use of X-ray imaging and analyze how the temporally resolved (15,136\,fps) thermal characteristics during solidification affect grain size and morphology (Figure\,\ref{fig:Overview}). A high-speed IR camera was used to capture the thermal radiation intensity maps near the melt pool during laser melting of C103 at different laser energies and scan speeds. The width of the melt pool was measured ex situ using scanning electron microscopy (SEM), and was used to identify the width of the melt pool boundary in the IR images. The liquidus temperature was then assigned to the melt pool boundary in the width direction in IR images. This facilitates the conversion of IR maps to temperature maps assuming a constant emissivity around the melt pool and by integrating spectral radiation only across the wavelength range detected by the camera, as opposed to the Stefan-Boltzmann scaling applied in previous literature. Using this methodology, the melt pool shape is identified and its moving velocity is analyzed. Thermal characteristics of the solidification process including temperature gradients ($G$), cooling rates ($dT/dt$), and solidification velocities ($R$) were analyzed. Ex situ SEM images reveal two distinct solidification modes, including epitaxial growth through the melt pool boundary and nucleated equiaxed grains within the melt pool. These microstructures are analyzed with respect to local solidification velocities and thermal gradients. This work demonstrates a methodology to resolve detailed time-dependent thermal characteristics during solidification and provides insights into the link between thermal process conditions and solidification behavior during metal AM.

\section{Methods}
\label{sec:method}

\subsection{Single-track laser scanning: experimental setup and procedure}
\label{sec:setup}

\linespread{1}

Single-track laser experiments were performed in a custom inert laser processing chamber interfaced with a high-speed IR camera (see Supporting Information S1). The chamber offers the basic functionality of a conventional L-PBF 3D printer for conducting rapid solidification experiments and was modified from an experimental setup in previous work by Mullin et al.\cite{mullin2024rapid}. The employed heat source is an IPG Photonics 400\,W Nd:YAG continuous laser operating at a wavelength of 1070\,nm.  Optical components, including a collimator, mirror galvanometer, and focus lens, are integrated to collectively manipulate and focus the laser beam. A customized Python code automates single-track laser scanning experiments by controlling both the galvanometers and the laser. The beam is focused down to $\approx$\,80\,$\mu$m on the build plate, which is situated on a motorized linear stage in the chamber.

A high-speed infrared camera (Telops M3K) equipped with a 1$\times$ lens (Telops) is positioned on the left side of the chamber, enabling the recording of the emitted thermal radiation within 3--5.5\,$\mu$m wavelength range (the wavelength range of the camera detector) during laser scanning.  Thermal radiation from the melt pool is reflected by a gold mirror (Thorlabs, 1\," Protected Gold Mirror, which has more than 96\,$\%$ reflectivity between 3--5.5\,$\mu$m) and passes through a Ge window (transparent between 3--5.5\,$\mu$m,  see Supporting Information S2)  to be captured by the IR camera with a viewing angle of $\theta\,\approx\,\,$60$^{\circ}$ as shown in Figure\,S1(a). $S_\text{effect,y}\!=$\,S/sin\,$\theta$ is the effective pixel size along the y-axis with a viewing angle of $\theta$. Here, $S\!=$\,30\,$\mu$m is the spatial resolution or the pixel size of the IR camera.  Therefore, $S_\text{effect,y}\,\approx\,34\,\mu$m, and $S_\text{effect,x}$ (the effective pixel size along the x-axis) is the same as the spatial resolution of the IR camera which is 30\,$\mu$m.

After placing the polished C103 sample on the build plate, the chamber is purged with ultra-high purity argon ($>$99.99\,$\%$) at a rate of 30 liters per hour.  An oxygen sensor is integrated with the argon outlet exhaust to monitor oxygen levels in the chamber. Once the oxygen levels are below 5\,ppm, the laser scanning process is performed on C103 alloy under different laser scanning powers (130, 170\,W) and scanning speeds (200, 400, 600\,mm/s). The scanning is captured with the high-speed infrared camera with a 128\,px$\times$64\,px (3.84\,mm$\times$2.18\,mm) window size, 20\,$\mu$s exposure time, and 15,136\, fps frame rate. The maximum frame rate of the IR camera is 100,000 fps but requires a window size too small to resolve melt pool temperature fields (64\,px$\times$4\,px). The selected frame rate is sufficiently fast while allowing a larger window size to image the melt pool and the laser track. The received signal via the IR camera is in the form of a digital signal (Non-Uniformity Correction, NUC value), necessitating an accurate conversion of the IR intensity maps into temperature fields before further thermal analysis.

\subsection{Temperature conversion}
\label{sec:boundary intensity}

\linespread{1}
\begin{figure*}
\centering
\includegraphics[width=.95\textwidth]{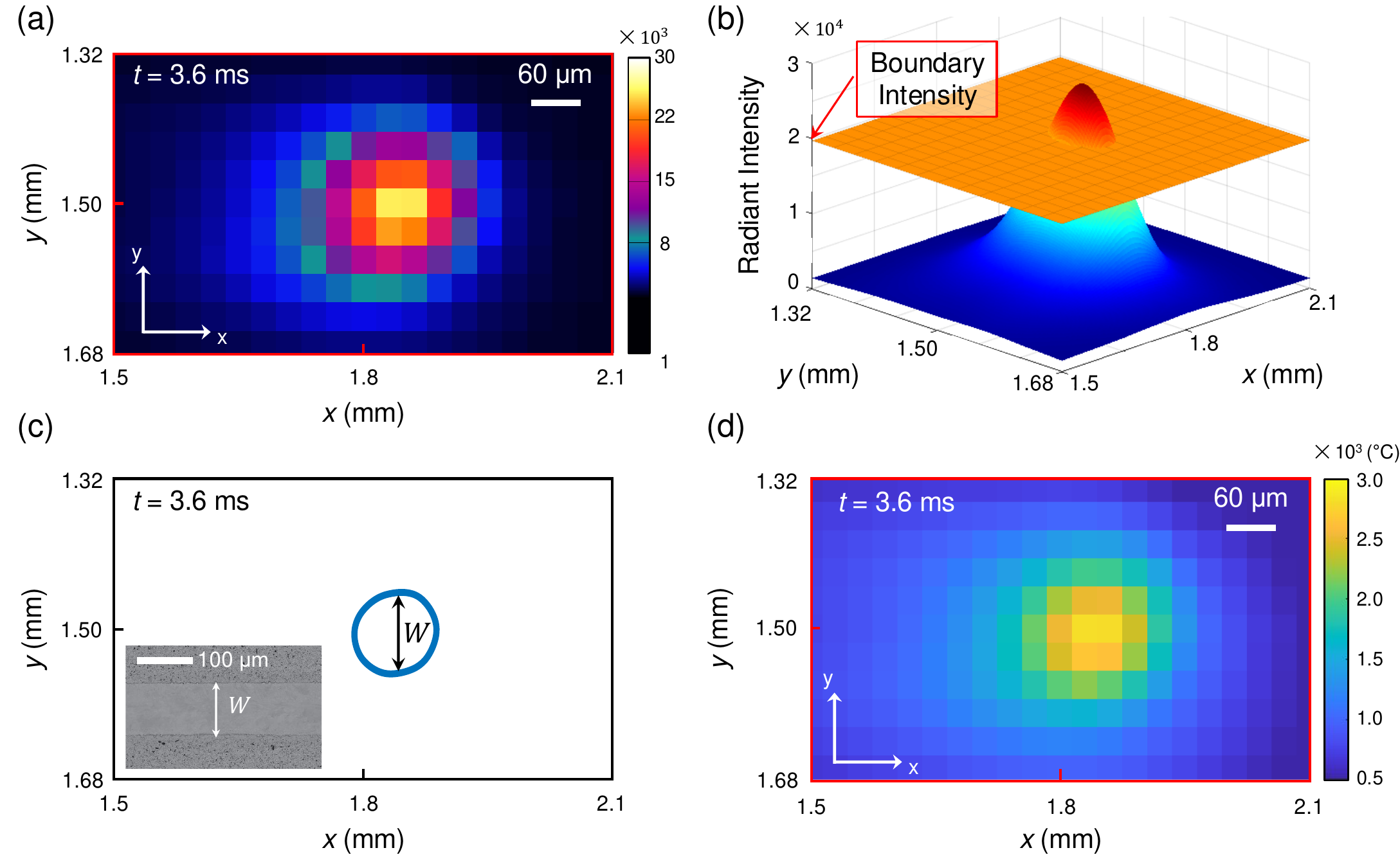}
\caption{Conversion of measured intensities (NUC) from the IR camera to temperature. (a) An example of a raw IR image of the melt pool of C103 showing measured IR intensity (NUC) in each pixel (30 $\times 34 \mu$m) under the laser scanning condition of 170\,W and 600\,mm/s. The same raw IR intensity map is replotted in (b), where the vertical axis is the magnitude of the NUC value and a smooth function is applied to the curved surface to reduce the coarsening effect due to the low spatial resolution of the IR camera. A horizontal plane is superimposed on this 2D map to identify the melt pool boundary. (c) The melt pool shape extracted from the intersection between the horizontal plane and the intensity map in (b), where the width ($W$) of the melt pool agrees with the width ($W$) shown in ex situ SEM images (inset). (d) The temperature map around the melt pool converted from the raw IR image. }
\label{fig:conversion}
\end{figure*}

Obtaining accurate temperature maps from IR measurements requires careful calibration because the detected IR signal depends on the spectral emissivity of the sample and signal attenuation through optical windows and mirrors. These are challenging and time consuming to measure at high temperatures, particularly near the liquidus of high temperature alloys like C103 ($T_\text{liquidus}\!\approx$\,2411\,$^\circ$C). Therefore, recent works have assigned the liquidus temperature to the melt pool boundary identified through synchronous X-ray imaging \cite{gould2021situ}. The temperatures of other pixels in the image were scaled based on the Stefan-Boltzmann law ($E\!\sim\!T^4 $) which integrates thermal radiation across the entire spectrum with a gray body assumption. However, since almost all IR cameras only detect radiation within a narrow wavelength band, the Stephan-Boltzmann approach will lead to an inaccurate scaling relationship. In this work, the melt pool boundary is first identified without X-ray imaging by simply identifying the width of the melt pool through ex situ SEM imaging. More importantly, we integrate thermal radiation only within the spectral range of the IR camera (Telops M3K) sensor (3-5.5 \,$\mu$m), when scaling the temperature map from the liquidus temperature.  Figure\,\ref{fig:conversion}(a) shows the measured raw IR intensities (NUC value) for a laser scanning condition of 170\,W and 600\,mm/s at a specific time ($t\!=\,$3.6\,ms) where $x$-direction is the laser scanning direction (from left to right). The same raw IR intensity map is replotted as shown in Figure\,\ref{fig:conversion}(b) where the vertical axis is the magnitude of the NUC value and a smooth function (spline in MATLAB) is applied to the curved surface to reduce the coarsening effect due to the low spatial resolution of the IR camera. Since the front half of the melt pool boundary should technically be around the liquidus temperature assuming very small superheating during melting, an arbitrary horizontal plane intersects this curved surface to form an imaginary "melt pool boundary" where NUC values on the melt pool boundary are all the same representing the same temperature. The vertical position of the horizontal plane is then adjusted until the resulting width of the imaginary melt pool in the $y$-direction ($W$ in Figure\,\ref{fig:conversion}(c)) agrees with the actual width of the melt pool measured through ex situ SEM imaging (inset of Figure\,\ref{fig:conversion}(c)). The NUC value or the raw intensity value thus identified (i.e., the final vertical position of the horizontal plane) represents the signal detected by the IR camera corresponding to the liquidus temperature ($T_\text{liquidus}\!\approx$\,2411\,$^\circ$C) of the sample \cite{mullin2024rapid}, which already accounts for attenuation through optical parts in the experimental setup. In the example shown in Figure\,\ref{fig:conversion}(a-c), this critical IR intensity value is identified to be 19720 which is the critical NUC value under the selected camera acquisition settings (20\,$\mu$s exposure time, and 15,136\,fps frame rate). Here, the liquidus temperature ($T_\text{liquidus}$) is used to correlate with the melt pool boundary intensity instead of the solidus temperature because the width of the laser track is determined by the melting process.

To further obtain the temperature map, we first numerically integrate Planck’s law, which describes blackbody radiation, over the 3--5.5 $\mu$m wavelength range, to derive the theoretical emitted energy within this band from the sample surface (Supporting Information Figure\,S3(a) blue line). The In-Band Irradiance (IBI), which is the blackbody radiation striking the camera detector's surface with a unit of W/m$^2$, is only a portion of the emitted blackbody radiation due to a view factor (Supporting Information Figure\,S3(a) red line). The view factor is dependent on the camera viewing angle and distance to the sample which is determined to be 1/8 for our experimental condition (Supporting Information Figure\,S3(b)). A 9th-order polynomial function is fitted to describe the numerical relationship between the IBI and temperature (Supporting Information Figure\,S3(b)). This relationship roughly describes how the detected IR raw signal scales with temperature for a blackbody, which differs from the $E\!\sim\!T^4 $ relationship in the Stefan-Boltzmann Law. The non-unity in-band emissivity of the sample, non-ideal spectral absorbance of the sensor, and transmittance of the optics in the IR camera require further correction to convert the theoretical IR signal (IBI) to the actual detected IR signal (NUC). To address this, the camera software (Reveal IR) offers a built-in correction to convert the IBI to NUC, which is also influenced by the camera acquisition setting, such as exposure time and frame rate. Specifically, for our experimental camera setting, as explained in Section\,\ref{sec:setup}, the relationship between NUC and IBI is introduced below.

\begin{equation}
\text{NUC} = \varepsilon \times 21.045 \times \text{IBI} + 1371.2
\label{NUC_IBI}
\end{equation} 

where  $\varepsilon$ is the effective emissivity of the C103 alloy, which accounts for the intrinsic in-band emissivity of C103 and additional signal attenuation from the gold mirror and the Ge window. The final calibration curve between the NUC and temperature, obtained using this conversion method, is shown in Figure\,S3(c) (Supporting Information, Section\,S3). Based on this NUC vs Tempearure relationship, the effective emissivity of C103 around the melt pool boundary was determined by adjusting its value until the NUC equals the critical NUC value (19720 in Section\,\ref{sec:boundary intensity}) when the temperature is the liquidus temperature of C103 metal. The effective emissivity of C103 at the liquidus temperature thus determined is $\varepsilon$ = 0.083. The total hemispherical emissivity averaged over all wavelengths and hemispherical directions of C103 reported in the literature is 0.28 at 800\,$^{\circ}$C and 0.4 at 1200\,$^{\circ}$C \cite{satya2017niobium}. While the emissivity of C103 in its liquid state is not reported in previous studies, it is generally observed that the emissivity of metals in their liquid state can be lower than in their solid state \cite{cagran2009normal}. In addition, emissivity values are sensitive to surface conditions and smooth and non-oxidized surfaces have lower emissivity than rough and oxidized counterparts. Given that the effective emissivity reported in this work is only the in-band emissivity between 3--5.5 $\mu$m in the camera viewing direction at the liquid-solid transition temperature for a polished C103, and further accounts for attenuation via the gold mirror and Ge window, this value of 0.083 is reasonable.

As a result, this conversion process yields the temperature map of the entire domain as illustrated in Figure\,\ref{fig:conversion}(d). To obtain a more accurate map, the temperature-dependent emissivity of the molten liquid metal and the solid C103 needs to be measured, which is beyond the scope of this work. However, in this work, since we are mainly interested in characterizing the temperature gradient and the cooling rate at the melt pool boundary, the assumption of a constant emissivity ($\varepsilon\!\approx$0.083) in calculating the temperature near the melt pool boundary is appropriate.

\section{Results}

\subsection{Ex situ microstructure characterization using SEM}
\label{sec:ex-situ}

\linespread{1}
\begin{figure*}
\centering
\includegraphics[width=0.7\textwidth]{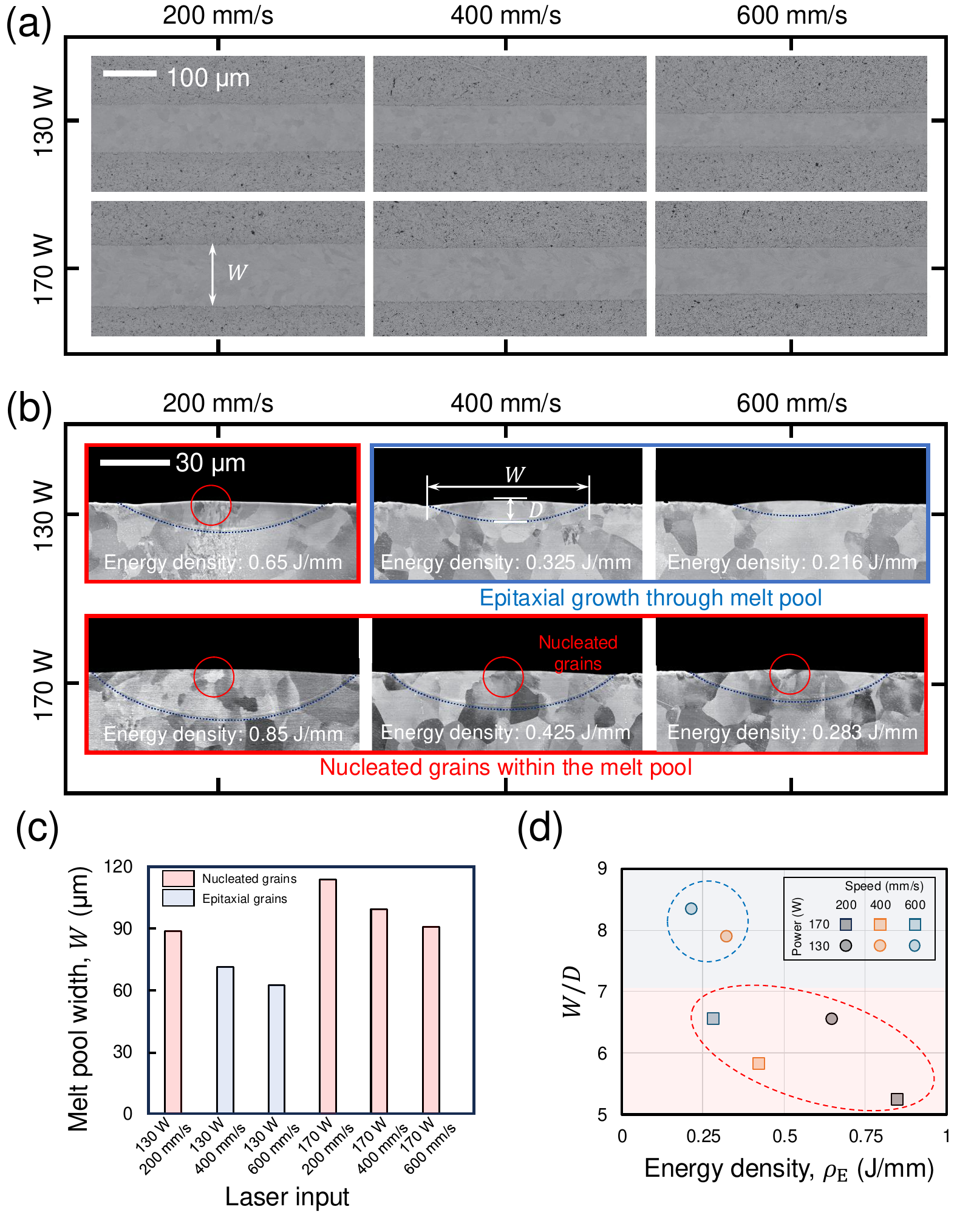}
\caption{Ex situ characterization of the microstructure and dimensions of laser melted C103. (a) SEM images of the top-down view of the printed material after laser scanning. (b) SEM images of the cross-section of the melt pool.  (c)  The measured width of the melt pool corresponds to the width of the melt pool ($W$) in (a) under each laser parameter. The red bar graph shows the nucleated grains within the melt pool ($W\,\gtrapprox$\,90\,$\mu$m), while the blue bar graph shows the epitaxial growth through the melt pool ($W\,\lessapprox$\,90\,$\mu$m). (d) The aspect ratios ($W/D$) of the melt pool are compared with the energy density ($\rho_\text{E}$) of the input laser. Higher aspect ratio values ($\gtrsim$\,7.8) indicate nucleated grains, while lower aspect ratios ($\lesssim$\,7) indicate epitaxial growth.}
\label{fig:exsitu}
\end{figure*}

Figure\,\ref{fig:exsitu} shows micrographs of a single-track experiment conducted on a wrought C103 substrate. After scanning the C103 alloy under different laser conditions ($\rho_\text{E} \approx$  0.2 -- 0.8\,J/mm, where $\rho_\text{E}$, the energy density, is the ratio between the laser power and the laser scan speed), the top-down view (Figure\,\ref{fig:exsitu}(a)) and the cross-sectional view (Figure\,\ref{fig:exsitu}(b)) of the samples are imaged using a scanning electron microscope (ThermoFisher, Apreo C) with a backscatter (BSE) detector. The measured widths ($W$) of the scanned melt pool extracted from Figure\,\ref{fig:exsitu}(a) are plotted in Figure\,\ref{fig:exsitu}(c). The cross-sectional view of the laser melted C103 in Figure\,\ref{fig:exsitu}(b) shows the width ($W$), the depth ($D$), and the microstructure of the melt pool. The more detailed SEM images can be found in Supporting Information S4. An intriguing observation is that track melts with a wider melt pool width ($\gtrsim$\,90\,$\mu$m) shows nucleated grains within the melt pool, as indicated by the red circle in Figure\,\ref{fig:exsitu}(b). In contrast, narrower melt pool width ($\lesssim$\,90\,$\mu$m) resulted in only epitaxial growth from the melt pool boundary, under the few experimental conditions tested.    Figure\,\ref{fig:exsitu}(d) illustrates the aspect ratio of the melt pool ($W/D$) alongside the 2D energy density ($\rho_\text{E}$) of the scanning laser for the six laser parameters examined. Relatively lower aspect ratio values ($W/D$\,$\lesssim$\,7) indicate nucleated grains, while relatively higher aspect ratios ($W/D$\,$\gtrsim$\,7.8) indicate epitaxial growth where the transition is at an aspect ratio of approximately 7--8.   Despite the observed trend between melt pool width, aspect ratio, and grain growth behavior, the thermal boundary conditions at which microstructural changes occur have yet to be determined for C103 and, more broadly, for refractory alloys. This is primarily due to the challenges associated with characterizing these values at high melting points in an inert environment. A detailed characterization of spatial and temporal temperature gradients, cooling rates, and solidification velocities during the experiment may shed insights into a more universal and machine-agnostic understanding of processing-microstructure relationships.

\linespread{1}
\begin{figure*}
\centering
\includegraphics[width=0.7\textwidth]{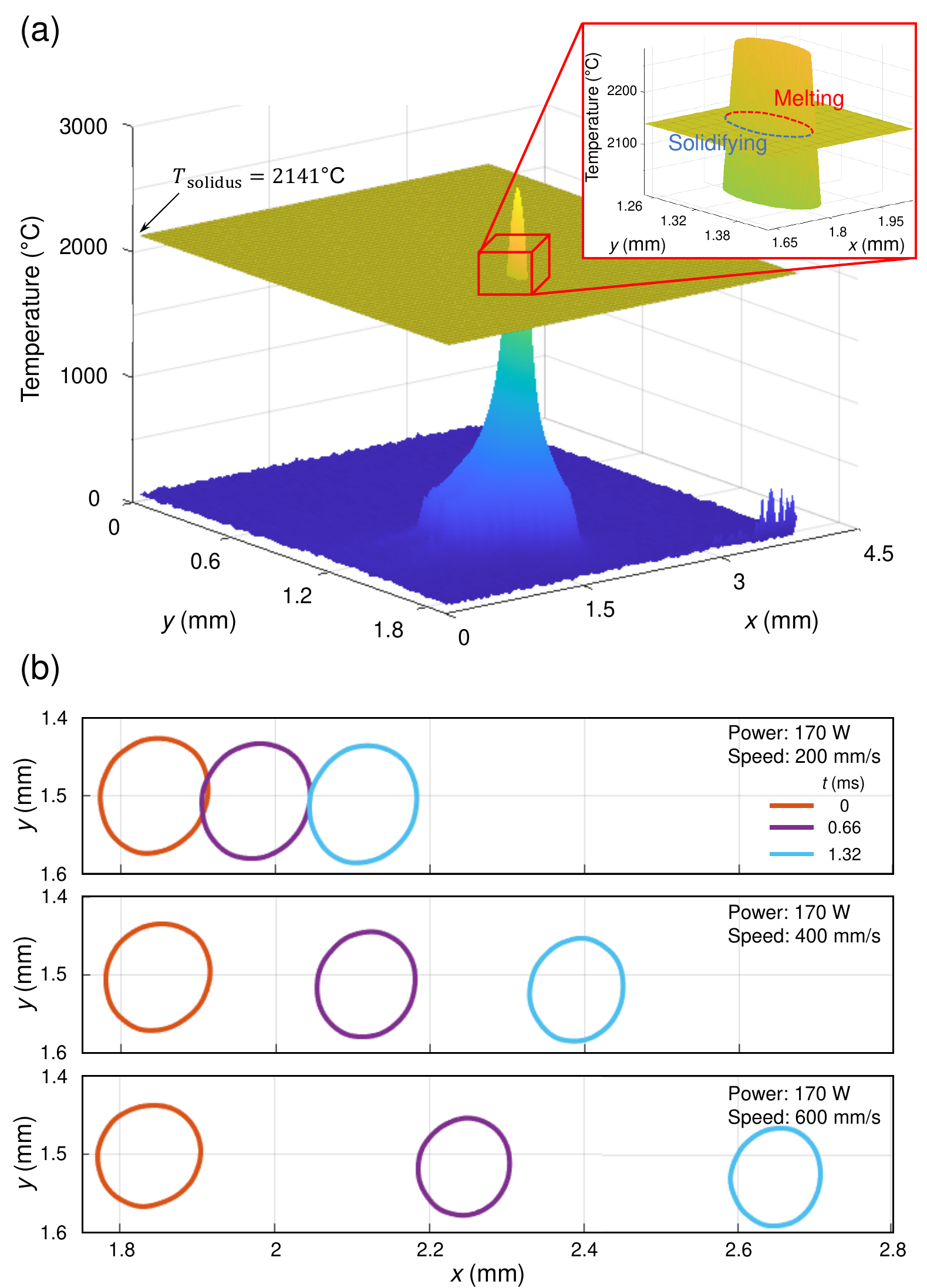}
\caption{Melt pool boundary and its movement over time. (a) 2D temperature map under a laser scanning parameter of 170\,W and 600\,mm/s. The horizontal plane is the solidus temperature of the C103 alloy. The inset shows the magnified view of the melt pool.  The circular line (red and blue dotted line) that intersects the 2D temperature map and the solidus temperature of C103 is the boundary of the melt pool. (b) the melt pool boundary at multiple time steps under various laser scan speeds.}
\label{fig:boundary}
\end{figure*}

\subsection{The melt pool shape and velocity}\label{sec:temp map}

The time-dependent temperature maps can be further processed to extract the temperature gradient ($G$), the heating/cooling rate ($dT/dt$), the 2D melt pool shape, and the solidification velocity ($R$) along the melt pool boundary, which are all important thermal parameters potentially affecting grain growth \cite{shao2019grain,farshidianfar2016effect,thampy2020subsurface,prasad2020towards}. Due to the relatively coarse pixel size of the IR camera (30 $\mu$m and 34 $\mu$m in $x$ and $y$ directions, respectively), the 2D temperature map is first interpolated to form a smoother map (using spline in MATLAB), as shown in Figure\,\ref{fig:boundary}(a) where the interpolated temperature map was plotted against $x$ and $y$ coordinates. To locate the melt pool boundary, especially the rear half of the melt pool where solidification occurs, a horizontal plane at the solidus temperature of C103 ($T_\text{solidus}\!\approx$\,2141\,$^\circ$C) was constructed to intersect the interpolated temperature surface to form a closed-loop line that represents the melt pool boundary, as shown in the inset of Figure\,\ref{fig:boundary}(a). More strictly speaking, it is more appropriate to use the liquidus temperature ($T_\text{liquidus}\!\approx$\,2411\,$^\circ$C) to identify the front half (melting, as shown in the red dotted line in the inset of Figure\,\ref{fig:boundary}(a)) of the melt pool and the width of the melt pool, which was employed in the earlier Section\,\ref{sec:boundary intensity}.  Since grain growth may occur during solidification,  solidus temperature is used for a more accurate depiction of the rear half of the melt pool shape. Here, we also assumed a small subcooling temperature compared to the thermodynamic solidus temperature. This assumption needs to be further investigated due to the high cooling rates employed in the study. Figure\,\ref{fig:boundary}(b) shows the melt pool boundary at multiple time steps (0, 0.66, 1.32\,ms) under various laser scan speeds (200, 400, 600\,mm/s) for a laser power of 170\,W. As expected, a lower scan speed (200\,mm/s) results in a slightly larger melt pool size compared to a higher scan speed (600\,mm/s), presumably due to a higher energy density (for a more detailed melt pool shape comparison, see Supporting Information Section\,S5.). The velocities at which the melt pools move extracted from Figure\,\ref{fig:boundary}(b) agree well with the laser scanning velocities. The melt pool shape and width remain relatively constant, suggesting that the system is approximately approaching the steady state  (see Supporting Information Section\,S6).

\linespread{1}
\begin{figure*}
\centering
\includegraphics[width=0.8\textwidth]{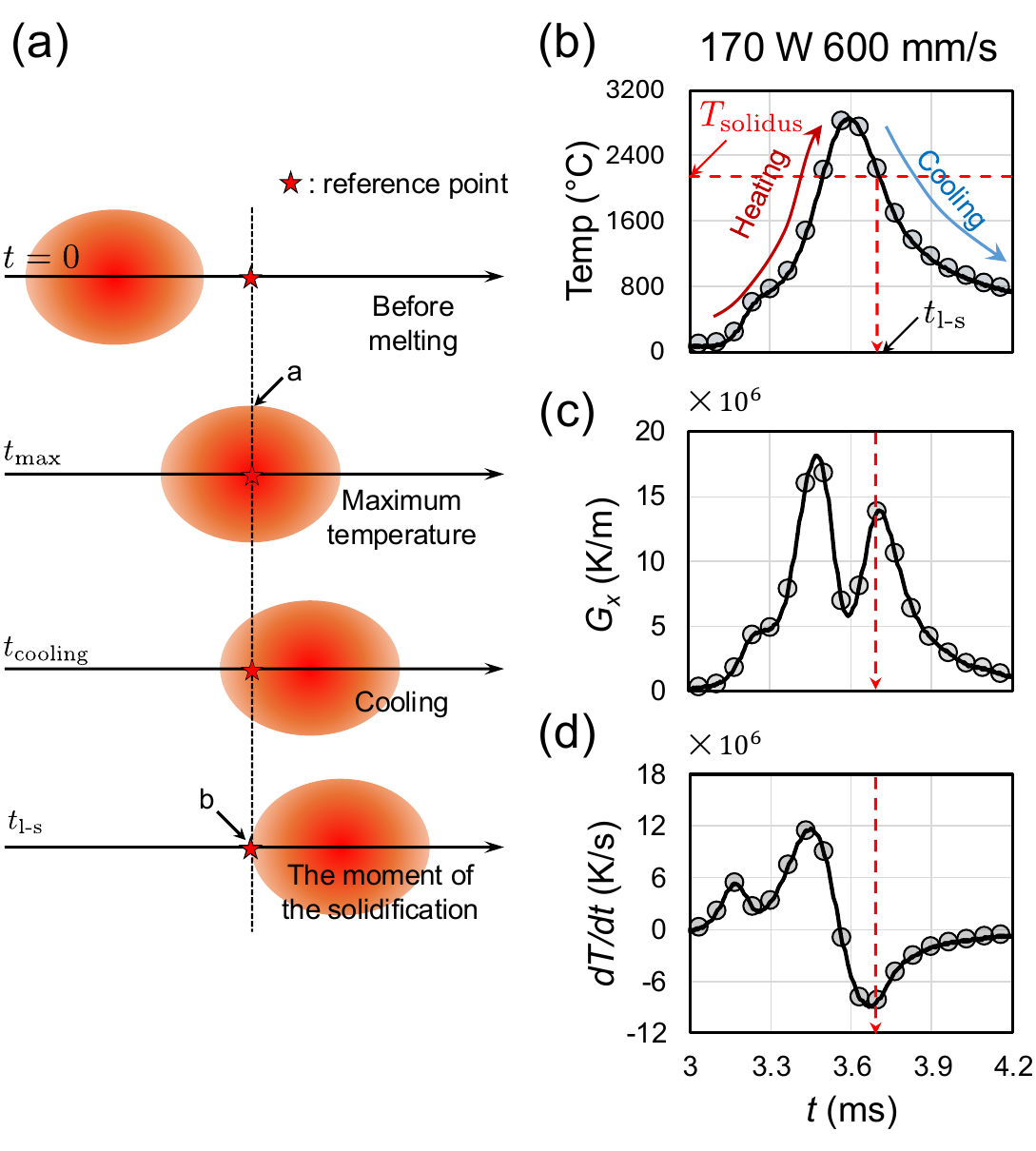}
\caption{Temporal thermal characteristics along the laser scanning direction. (a) A schematic illustrates the movement of the melt pool during laser scanning in the $x$-direction, with a star marking a reference point along the laser scanning line. (b) The temperature ($T$), (c) temperature gradient ($G_x = dT/dx$), and (d) heating/cooling rate ($dT/dt$) of a reference pixel along the laser scanning line over time are depicted. A horizontal dotted line in the temperature map (b) indicates the solidus temperature of the C103 alloy, while a vertical dotted line in (b),(c), and (d) marks the moment of solidification.}
 \label{fig:thermo temporal}
 \end{figure*}

\subsection{Thermal characterization: temporal analysis}
\label{sec:temp gradient_temp}

To characterize the temporal temperature response, the following analysis examines a single point along the laser scanning direction($x$-direction) as illustrated in Figure\,\ref{fig:thermo temporal}(a). The melt pool moves in the $x$-direction over a reference point denoted with a star. As the laser moves, this reference point melts and subsequently solidifies. The temperature ($T$), the temperature gradient ($G$), and the heating/cooling rate ($dT/dt$) of the reference point for a representative laser parameter (170\,W, 600\,mm/s) are shown in Figure\,\ref{fig:thermo temporal}(b-d). The circular data points represent the actual data captured by the infrared camera, while the smooth solid line is interpolated from the smoothed 2D temperature profile using the same method as described in Figure\,4a. The temperature response in Figure\,\ref{fig:thermo temporal}(b) shows rapid heating and cooling within only 0.6--1.2\,ms, highlighting the necessity for high-speed characterization which is achievable with our camera frame rate. The peak temperature in the melt pool could be underestimated due to an expected lower emissivity of liquid metal than the solid. In addition, the asymmetry of the temporal response of temperature, or more rapid heating than cooling, agrees well with previous thermal simulations and analytical solutions for a moving heat source on a solid substrate \cite{van2007solutions,forslund2019analytical}. Solidification occurs when the temperature drops to the solidus temperature (dashed horizontal line) during cooling, corresponding to $t=t_\textbf{l-s}$ in Figure\,\ref{fig:thermo temporal}(b). 

The temperature gradient on the $x$--$y$ plane is
 \begin{equation}
G = \sqrt{\left(\frac{dT}{dx}\right)^2+\left(\frac{dT}{dy}\right)^2},
\label{G}
\end{equation}

Note that the total temperature gradient involves the contribution of a temperature gradient in the $z$-direction, which is not captured by the IR camera. However, only the $x$-direction temperature gradient $G_x$ is needed along the center laser scan path due to symmetry in the $y$ direction,  

\begin{equation}
G_x = \frac{dT}{dx} \approx \frac{\Delta T_{x}}{\Delta x},
\label{G_x}
\end{equation}

where $\Delta x$ is the distance between adjacent points in the smoothed temperature map in the $x$-direction (3 $\mu$m after smoothing the temperature map), and $\Delta T_{x}$ is estimated using a linear approximation of the temperature centered difference method (for more details about the gradient derivation, see Supporting Information S7).

The temporal response of the temperature gradient ($G_x$) (Figure\,\ref{fig:thermo temporal}(c)) shows two peaks around 3.4\,ms and 3.7\,ms, which happen during melting and solidification (by identifying the temperatures at 3.4\,ms and 3.7\,ms in the temperature versus time curve). On the contrary, the temperature gradient within the melt pool is much lower, despite the highest local laser energy flux. This can result from a higher effective thermal conductivity of the liquid due to convective flow, effectively reducing the temperature gradient within the melt pool. In addition, the temperature gradient peak during melting is higher than the temperature gradient peak during solidification, which agrees with the asymmetry in the heating and cooling process. Figure\,\ref{fig:thermo temporal}(d)  also ($dT/dt$) shows more rapid heating ($\approx$ $12\,\times10^6\,$K/s) as compared to cooling ($\approx$ $7\,\times10^6$\,K/s). The magnitude of thermal gradient upon cooling is within estimated ranges for L-PBF (10$^4$ - 10$^8$ K/m) \cite{mullin2024cracking}. Though thermal gradients under laser melting conditions have not previously been measured for Nb-base or refractory alloys generally, thermal simulations have produced values within the range of 10$^7$ K/m for C103 \cite{miklas2022additive} and 10$^7$-10$^8$ for pure refractory metals with higher thermal conductivity (Ta, Mo) \cite{fernandez2021crystallographic,thijs2013strong}. It is anticipated that the lower effective thermal conductivity and higher absorptivity of powder feedstock for L-PBF may influence the thermal gradients, which demands further investigation.

\subsection{Thermal characterization: on the melt pool boundary}
\label{sec:temp gradient_spetial}

\linespread{1}
\begin{figure*}
\centering
\includegraphics[width=1\textwidth]{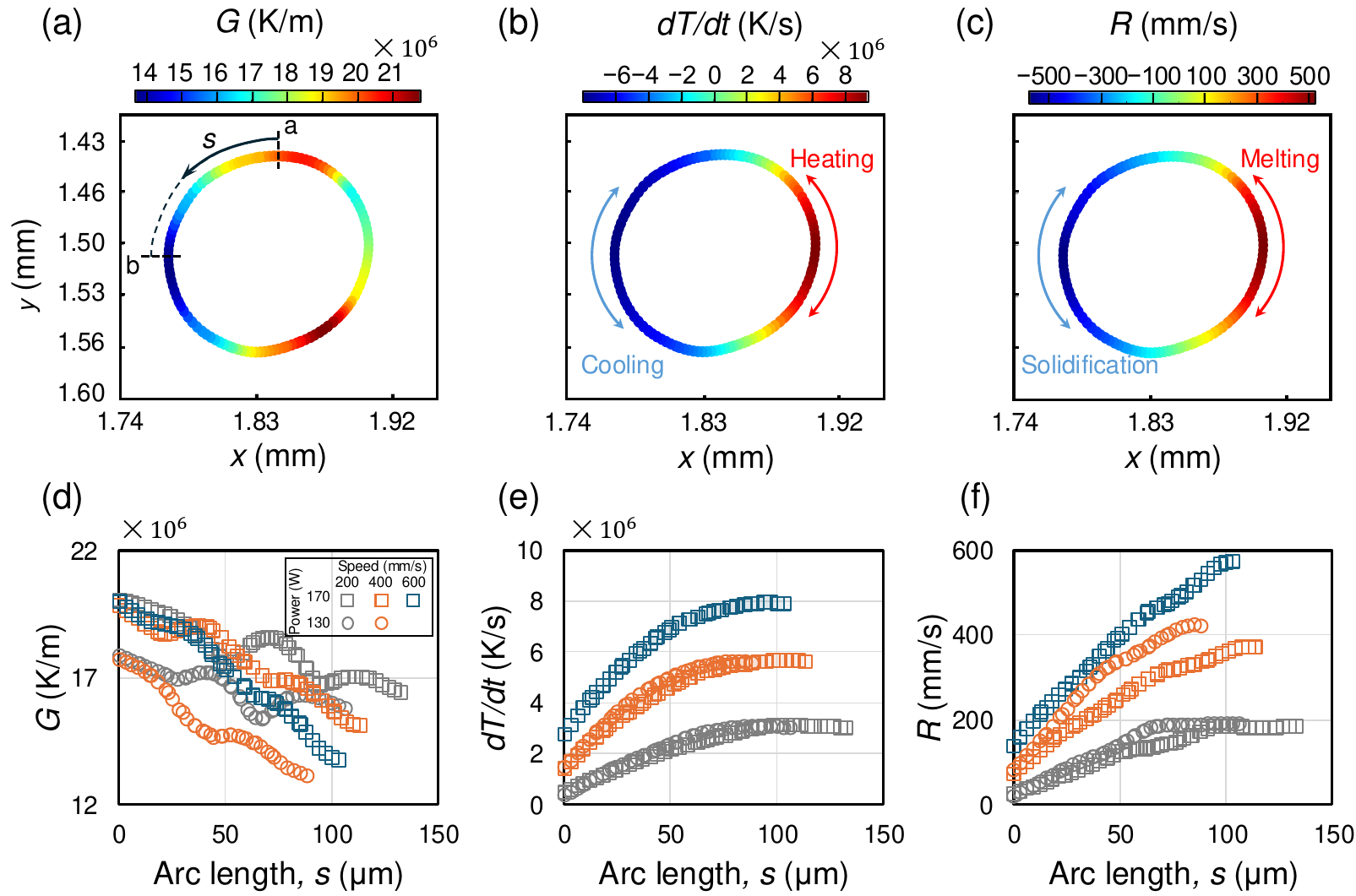}
\caption{Spatial thermal information along the melt pool boundary. (a) The temperature gradient ($G$), (b) cooling/heating rate ($dT/dt$), and (c) solidification velocity ($R$) for 170\,W 600\,mm/s laser scanning at the 3.6\,ms time frame. The laser is moving in the +$x$-direction. Temperature gradient and solidification velocity do not have a $z$-component contribution, as the spatial analysis is based on 2D temperature measurements. The thermal information along the arc from the outer edge to the end of the melt pool (as illustrated in (a)) is also extracted from the spatial thermal data, such as (d) the thermal gradient ($G$), (e) the cooling/heating rate ($dT/dt$), and (f) the solidification velocity ($R$), which are plotted against the arc length ($s$) in different laser scan conditions.}
 \label{fig:thermo spatial}
 \end{figure*}

In addition to investigating the temporal response of one point along the laser scanning line, we also characterize the spatial distribution of temperature gradient ($G$), heating/cooling rate ($dT/dt$), and solidification velocity ($R$) on the melt pool boundary, as shown in Figure\,\ref{fig:thermo spatial}(a-c). For a laser scanning condition of 170\,W and 600\,mm/s at a fixed time of 3.6\,ms, the melt pool boundary is extracted using the same approach as in Figure\,\ref{fig:boundary}. The temperature gradient ($G$) and solidification velocity ($R$) on the melt pool boundary do not account for a $z$-direction component, as the spatial analysis is based on 2D temperature measurements. The temperature gradient ($G$) was calculated based on Equation\,\ref{G}, which accounts for the gradients in the $x$ and $y$ directions. Its numerical form is

 \begin{equation}
G \approx \sqrt{\left(\frac{\Delta T_{x}}{\Delta x}\right)^2+\left(\frac{\Delta T_{y}}{\Delta y}\right)^2},
\label{G_spatial}
\end{equation}

where $\Delta x$ and $\Delta y$ represent the distance between adjacent points in the smoothed temperature map in the $x$ and $y$ directions (3\,$\mu$m and 3.4\,$\mu$m, respectively after smoothing the temperature map), and $\Delta T_{x}$ and $\Delta T_{y}$ are estimated using a linear approximation of the temperature centered difference method (see Supporting Information S7 for more details). As shown in Figure\,\ref{fig:thermo spatial}(a), the temperature gradient varies only slightly from $14\!\times\!10^{6}$\, to $21\!\times\!10^{6}$\, K/m on the melt pool boundary under this laser parameter. The top and bottom edges of the melt pool boundary exhibit the highest temperature gradient values, indicating that the temperature change is maximized at the outer width along the $y$-direction, which is lateral to the laser scanning direction. The heating/cooling rate ($dT/dt$) at 3.6\,ms is also plotted on the contour, as shown in Figure\,\ref{fig:thermo spatial}(b). Specifically, the front half of the melt pool boundary represents the heating section, where $dT/dt$ is positive. Conversely, the rear half of the melt pool boundary represents the cooling section, where $dT/dt$ is negative. The cooling rate is the highest at the trailing edge of the melt pool along the laser scanning path, where the thermal gradient is the lowest ($\approx\!-$8\,$\times10^{6}$\,K/s).

The solid-liquid interface velocity, $R$, also known as the solidification velocity, can be calculated based on the ratio between the time derivative of the cooling rate  ($dT/dt$) and the temperature gradient  ($G$), as shown in previous studies \cite{raplee2017thermographic}

\begin{equation}
R = \frac{\left(\frac{dT}{dt}\right)}{G},
\label{R}
\end{equation}

Similarly, the negative value of $R$ represents solidification, whereas the positive $R$ represents melting. Mathematically, the solidification velocity derived from this equation represents the solid-liquid interface velocity in the temperature gradient direction. In our work, since $G$ does not include a $z$-direction component from the 2D temperature measurement, the $R$ value shown in Figure\,\ref{fig:thermo spatial}(c) only represents the projected velocity on the $x$-$y$ plane.  In Figure\,\ref{fig:thermo spatial}(a-b), $G$, $dT/dt$ at the trailing edge of the melt pool along the laser scanning path agrees well with values shown in Figure\,\ref{fig:thermo temporal}(c-d), confirming that our analysis is consistent. 

Additionally, the temporal response of the solidification velocity at the melt pool edge can be obtained by tracking the melt pool boundary from the side edge to the center, perpendicular to the laser scanning direction. The solidification velocity in the $y$-direction as a function of time is presented in Supporting Information  Section\,S8. It can be observed that the $y$-direction solidification velocity also accelerates as the edge moves toward the center. Both the gradient-based solidification velocity (Figure\,\ref{fig:thermo spatial}(c)) and the y-direction solidification velocity (Section\,S8) illustrate the increase in solidification velocity from the edge (point a in Figure\,\ref{fig:thermo temporal}(a)) to the center(point b in Figure\,\ref{fig:thermo temporal}(a)). The terminal velocity of $R$, which is the velocity at the trailing edge, agrees well with the laser scanning velocity, indicating that the system is approximately at the steady state.

\subsection{Microstructural transition during solidification}
\label{sec:CET map}

 The spatially resolved thermal features in Figure\,\ref{fig:thermo spatial}(a-c) allow further analysis of the columnar-to-equiaxed transition (CET), observed in four of the laser scanning conditions. Solidification of a melt pool typically progresses from the melt pool boundary toward the center of the melt pool in three dimensions. However, since we can only measure temperature on the top $x$-$y$ plane with no $z$-component, the following analysis is focused on investigating the thermal features ($G$, $dT/dt$ and $R$) leading to the grain morphology transition observed on the top $x$-$y$ plane, specifically from the side edge (point a in Figure\,\ref{fig:thermo temporal}(a)) to the trailing edge (point b in Figure\,\ref{fig:thermo temporal}(a)) of the melt pool. To illustrate this transition, thermal features are extracted along the arc ($s$) of the melt pool boundary from the uppermost edge to the trailing edge, as illustrated from 'a' to 'b' in Figure\,\ref{fig:thermo spatial}(a). The corresponding temperature gradient ($G$), cooling rate ($dT/dt$), and solidification velocity ($R$) are then plotted against $s$ in different laser scanning conditions, as shown in Figure\,\ref{fig:thermo spatial}(d-f).

\begin{figure*}
\centering
\includegraphics[width=1\textwidth]{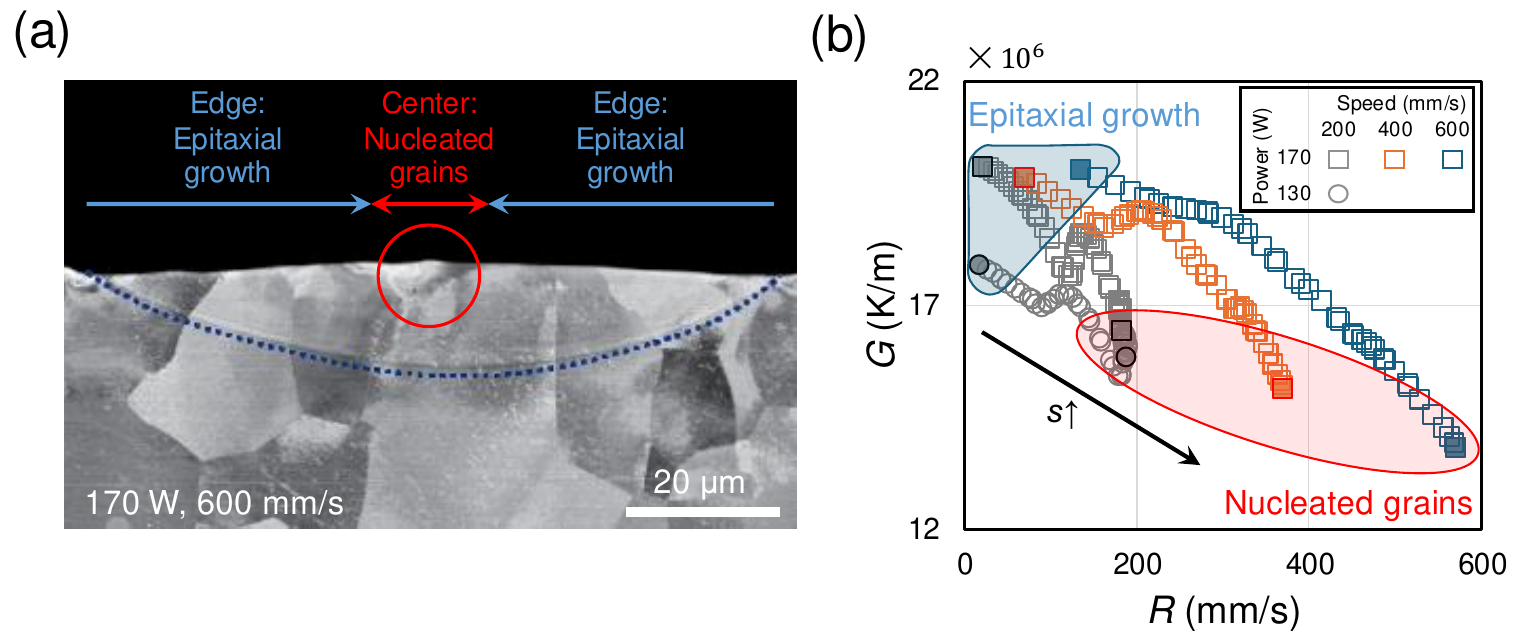}
\caption{Microstructure changes during the solidification. (a) shows SEM images of the cross-section of the melt pool under a laser scanning parameter of 170\,W and 600\,mm/s. The early stage of cooling exhibits epitaxial growth starting from the edge, while the later stage shows the formation of nucleated grains at the center of the melt pool. (b) represents the thermal phase map between temperature gradient and solidificiation velocity, illustrating the transition from epitaxial growth to nucleated grains. This transition occurs with a decreasing temperature gradient and increasing solidification velocity, consistent with previously reported CET maps \cite{kobryn2003microstructure,liu2019insight}.}
 \label{fig:phase map}
 \end{figure*}

 Figure\,\ref{fig:thermo spatial}(d) shows that the temperature gradient decreases from the edge to the center of the melt pool, while the cooling rate (Figure\,\ref{fig:thermo spatial}(e)) and solidification velocity (Figure\,\ref{fig:thermo spatial}(f)) increase from the edge to the center of the melt pool for all of the laser conditions. One interesting observation is that the cooling rate (Figure\,\ref{fig:thermo spatial}(e)) changes with the laser scan speed, and is effectively independent of the laser scan power. This result is consistent with findings in previous studies \cite{gong2014beam,chai2021effect}. This is because faster scan speeds reduce the time the laser spends on each location, limiting the material’s ability to absorb and retain heat, while the laser power primarily affects the size and temperature of the molten pool rather than the cooling rate. In particular, the faster scan speed (600 mm/s) results in a higher cooling rate. For the four laser conditions where a CET was observed (Figure\,\ref{fig:exsitu}(b)), epitaxial or columnar growth is observed at the edge, whereas nucleated or equiaxed grains are observed at the center of the melt pool track (see Figure\,\ref{fig:phase map}(a)). Correlating the spatially resolved thermal data with the location-dependent grain growth suggests that a higher temperature gradient, lower cooling rate, and lower solidification velocity on the edge of the melt pool more likely lead to epitaxial growth. The transition to equiaxed grains occurs as the temperature gradient decreases and the cooling rate and solidification velocity increase towards the center of the melt pool.  
%The temperature gradient ($G$) is shown in Figure\,\ref{fig:thermo spatial}(d) as a function of  arc length. Regardless of the laser scanning conditions, the general trend of the temperature gradient decreases as arc length increases. In other words, during the cooling process, the thermal gradient has its maximum value when cooling begins (at the edge) and continues to decrease, showing a minimum value at the end of the cooling process (at the center). Similarly, the cooling rate ($dT/dt$) and solidification velocity ($R$) are measured under different laser conditions. As shown in Figure\,\ref{fig:thermo spatial}(e), the cooling rate increases with increasing arc length. At the beginning of the cooling process, the cooling rate shows minimum values; however, at the end, it reaches its maximum value, which is in contrast to the trend of the thermal gradient. Interestingly, the increase in cooling rate follows the laser scan speed rather than the laser scan power, indicating that a higher scan speed results in a greater increase in cooling rate, thus determining the rate of cooling. The solidification velocity ($R$) can also be obtained, as shown in Figure\,\ref{fig:thermo spatial}(f). 
%Similar to the trend observed for the cooling rate, the solidification velocity ($R$) increases with increasing arc length  and is calculated from Equation \ref{R}, representing the same direction with the gradient direction.

\linespread{1}

 %As explained in Section\,\ref{sec:temp gradient_spetial}, the solidification process begins at the edge and progresses toward the center of the melt pool track. During this rapid cooling from edge to center, the microstructure could change, leading to increased porosity or defects that can impact the overall performance and reliability of the part. Under different laser scanning conditions, changes in microstructure have been observed. As illustrated in Figure\,\ref{fig:exsitu}(b), the relatively wider melt pool width ($\gtrsim$\,90\,$\mu$m) and lower aspect ratio values ($W/D$\,$\lesssim$\,7) indicate the presence of nucleated grains within the melt pool. However, these nucleated grains do not form at the onset of cooling; instead, the microstructure evolves throughout the cooling process. Figure\,\ref{fig:phase map}(a) presents a cross-sectional SEM image of the melt pool under 170\,W and 600\,mm/s laser conditions, which represent the case with nucleated grains on the melt pool. Solidification begins at the edge of the melt pool, where the long distance from the center results in lower heat input. As solidification progresses, the microstructure initially develops through epitaxial growth at the edge (indicated by the blue arrow regime) and transitions to nucleated grains at the top of the melt pool as cooling continues (indicated by the red arrow regime). In other words, the microstructure evolves from epitaxial growth to nucleated grains as it cools. 

The observed CET for C103 is consistent with other conventional metals, such as Ti–6Al–4V and IN718 \cite{kobryn2003microstructure,liu2019insight}, where the CET has been plotted on a temperature gradient ($G$) vs solidification velocity ($R$) maps. In previously reported CET maps between $G$ and $R$, the CET curve was calculated through simulation \cite{liu2019insight}, demonstrating that the transition occurs when the temperature gradient ($G$) decreases and the solidification velocity ($R$) increases during solidification. This understanding aligns with our observations for C103, as shown in Figure\,\ref{fig:phase map}(b). Specifically, Figure\,\ref{fig:phase map}(b) plots the temperature gradient ($G$) and solidification velocity ($R$) of all locations along the arc of the melt pool, extracted from Figure\,\ref{fig:thermo spatial}(a) and (c). Data in the top left region in Figure\,\ref{fig:phase map}(b) (blue shaded area) are located at the outer edge of the melt pool, and the bottom right region (red shaded area) shows data at the center (trailing edge) of the melt pool. The starting and ending points of the solidification are represented as filled data points, which are associated with epitaxial growth and nucleated grains, respectively. This approach provides experimental data to construct a CET map for C103, which provides guidance for microstructure control efforts during SLM.

\section{Conclusion}

In summary, in situ IR measurements of the Nb-based alloy C103 during laser melting were conducted to extract temporally resolved temperature fields during solidification. Microstructure development was examined ex situ, revealing epitaxial growth and nucleated grains under different laser conditions. To resolve the detailed thermal characteristics leading to the microstructures, we present a calibration methodology to convert the detected IR intensities to temperature maps by carefully considering in-band irradiance, view angle, non-ideality of optics as well as the emissivity of the sample. The melt pool boundary as well as its moving dynamics are presented. We analyze the transient response of a point along the laser scanning path, highlighting the rapid change of temperature, temperature gradients, and cooling rates within only 1 ms. In addition, spatially resolved temperature gradients, cooling rates, and solidification velocity on the melt pool boundary are presented, while the time-dependent solidification process from the melt pool edge inward is also illustrated. This approach enabled the creation of a thermal phase map that associates microstructural transition during the solidification process to the local temperature gradient and solidification rate. This study underscores the importance of in situ IR measurements for a detailed understanding of solidification conditions and microstructure development during additive manufacturing. Moreover, the ability to accurately measure temperature fields with high temporal resolution is crucial for addressing machine-to-machine variability in AM processes, which can lead to significant differences in microstructural outcomes. By enhancing our understanding of these thermal processes, the IR method offers a valuable tool for improving consistency and control across different AM systems.

Acknowledgment: The thermal measurement and analysis part of this work is supported by the National Science Foundation through the Materials Research Science and
Engineering Center (MRSEC) at UC Santa Barbara: NSF DMR–2308708 (Seed). The laser melting experimental setup and materials characterization are based upon work supported by the Department of Energy, National Nuclear Security Administration under Award Number(s) DE-NA0004152. O.A. Wander and K.M. Mullin also acknowledge support from the National Science Foundation Graduate Research Fellowship under Grant No. 2139319. We thank Prof. Patrick Hopkins and Hunter Schonfeld at the University of Virginia for discussions on emissivity measurement. We thank Wes Autran, Alex Côté, Vince Morton, Andrew Niehaus, and Hector Moreno from Telops for detailed discussions and suggestions on the temperature conversion process. We thank Alden Wong and Sachin Suri for their contributions to data analysis and development of experimental hardware. We thank S. Farzad Ahmadi for the comments on the figures. The research reported here made use of the shared facilities of the Materials Research Science and Engineering Center (MRSEC) at UC Santa Barbara: NSF DMR–2308708. The UC Santa Barbara MRSEC is a member of the Materials Research Facilities Network (www.mrfn.org).

Disclaimer: This report was prepared as an account of work sponsored by an agency of the United States Government. Neither the United States Government nor any agency thereof, nor any of their employees, makes any warranty, express or implied, or assumes any legal liability or responsibility for the accuracy, completeness, or usefulness of any information, apparatus, product, or process disclosed, or represents that its use would not infringe privately owned rights. Reference herein to any specific commercial product, process, or service by trade name, trademark, manufacturer, or otherwise does not necessarily constitute or imply its endorsement, recommendation, or favoring by the United States Government or any agency thereof. The views and opinions of authors expressed herein do not necessarily state or reflect those of the United States Government or any agency thereof.

$^{\dagger}$ These authors contributed equally to this work.
$^{*}$ Corresponding authors. E-mail addresses: tresap@ucsb.edu, yangying@ucsb.edu

%\appendix

%\section{Sample Appendix Section}

% \bibliographystyle{elsarticle-num} 
% \bibliography{cas-refs}
%\bibliographystyle{elsarticle-harv}
\bibliographystyle{unsrt}
\bibliography{NIFI_references}

\begin{thebibliography}{10}

\bibitem{kotadia2021review}
HR~Kotadia, G~Gibbons, Amit Das, and PD~Howes.
\newblock A review of laser powder bed fusion additive manufacturing of aluminium alloys: Microstructure and properties.
\newblock {\em Additive Manufacturing}, 46:102155, 2021.

\bibitem{raghavan2017localized}
Narendran Raghavan, Srdjan Simunovic, Ryan Dehoff, Alex Plotkowski, John Turner, Michael Kirka, and Suresh Babu.
\newblock Localized melt-scan strategy for site specific control of grain size and primary dendrite arm spacing in electron beam additive manufacturing.
\newblock {\em Acta Materialia}, 140:375--387, 2017.

\bibitem{dehoff2015site}
Ryan~R Dehoff, MM~Kirka, WJ~Sames, H~Bilheux, AS~Tremsin, LE~Lowe, and SS~Babu.
\newblock Site specific control of crystallographic grain orientation thrfough electron beam additive manufacturing.
\newblock {\em Materials Science and Technology}, 31(8):931--938, 2015.

\bibitem{david1989correlation}
SA~David and JM~Vitek.
\newblock Correlation between solidification parameters and weld microstructures.
\newblock {\em International materials reviews}, 34(1):213--245, 1989.

\bibitem{collins2016microstructural}
PC~Collins, DA~Brice, P~Samimi, I~Ghamarian, and HL~Fraser.
\newblock Microstructural control of additively manufactured metallic materials.
\newblock {\em Annual Review of Materials Research}, 46:63--91, 2016.

\bibitem{debroy2018additive}
Tarasankar DebRoy, Huiliang~L Wei, James~S Zuback, Tuhin Mukherjee, John~W Elmer, John~O Milewski, Allison~Michelle Beese, A~de Wilson-Heid, Amitava De, and Wei Zhang.
\newblock Additive manufacturing of metallic components--process, structure and properties.
\newblock {\em Progress in Materials Science}, 92:112--224, 2018.

\bibitem{das2003physical}
Suman Das.
\newblock Physical aspects of process control in selective laser sintering of metals.
\newblock {\em Advanced engineering materials}, 5(10):701--711, 2003.

\bibitem{echeta2020review}
Ifeanyichukwu Echeta, Xiaobing Feng, Ben Dutton, Richard Leach, and Samanta Piano.
\newblock Review of defects in lattice structures manufactured by powder bed fusion.
\newblock {\em The International Journal of Advanced Manufacturing Technology}, 106:2649--2668, 2020.

\bibitem{kimura2017effect}
Takahiro Kimura, Takayuki Nakamoto, Masataka Mizuno, and Hideki Araki.
\newblock Effect of silicon content on densification, mechanical and thermal properties of al-xsi binary alloys fabricated using selective laser melting.
\newblock {\em Materials Science and Engineering: A}, 682:593--602, 2017.

\bibitem{attar2014manufacture}
Hooyar Attar, Mariana Calin, LC~Zhang, Sergio Scudino, and J{\"u}rgen Eckert.
\newblock Manufacture by selective laser melting and mechanical behavior of commercially pure titanium.
\newblock {\em Materials Science and Engineering: A}, 593:170--177, 2014.

\bibitem{scime2019melt}
Luke Scime and Jack Beuth.
\newblock Melt pool geometry and morphology variability for the inconel 718 alloy in a laser powder bed fusion additive manufacturing process, addit. manuf. 29 (2019) 100830.
\newblock {\em J. ADDMA}, 2019.

\bibitem{cunningham2019keyhole}
Ross Cunningham, Cang Zhao, Niranjan Parab, Christopher Kantzos, Joseph Pauza, Kamel Fezzaa, Tao Sun, and Anthony~D Rollett.
\newblock Keyhole threshold and morphology in laser melting revealed by ultrahigh-speed x-ray imaging.
\newblock {\em Science}, 363(6429):849--852, 2019.

\bibitem{zhao2020critical}
Cang Zhao, Niranjan~D Parab, Xuxiao Li, Kamel Fezzaa, Wenda Tan, Anthony~D Rollett, and Tao Sun.
\newblock Critical instability at moving keyhole tip generates porosity in laser melting.
\newblock {\em Science}, 370(6520):1080--1086, 2020.

\bibitem{ren2023machine}
Zhongshu Ren, Lin Gao, Samuel~J Clark, Kamel Fezzaa, Pavel Shevchenko, Ann Choi, Wes Everhart, Anthony~D Rollett, Lianyi Chen, and Tao Sun.
\newblock Machine learning--aided real-time detection of keyhole pore generation in laser powder bed fusion.
\newblock {\em Science}, 379(6627):89--94, 2023.

\bibitem{oliveira2020processing}
Jo{\~a}o~Pedro Oliveira, AD~LaLonde, and J~Ma.
\newblock Processing parameters in laser powder bed fusion metal additive manufacturing.
\newblock {\em Materials \& Design}, 193:108762, 2020.

\bibitem{lewandowski2016metal}
John~J Lewandowski and Mohsen Seifi.
\newblock Metal additive manufacturing: a review of mechanical properties.
\newblock {\em Annual review of materials research}, 46:151--186, 2016.

\bibitem{hanzl2015influence}
Pavel Hanzl, Miroslav Zetek, Tom{\'a}{\v{s}} Bak{\v{s}}a, and Tom{\'a}{\v{s}} Kroupa.
\newblock The influence of processing parameters on the mechanical properties of slm parts.
\newblock {\em Procedia Engineering}, 100:1405--1413, 2015.

\bibitem{markl2016multiscale}
Matthias Markl and Carolin K{\"o}rner.
\newblock Multiscale modeling of powder bed--based additive manufacturing.
\newblock {\em Annual Review of Materials Research}, 46(1):93--123, 2016.

\bibitem{shi2020microstructural}
Rongpei Shi, Saad~A Khairallah, Tien~T Roehling, Tae~Wook Heo, Joseph~T McKeown, and Manyalibo~J Matthews.
\newblock Microstructural control in metal laser powder bed fusion additive manufacturing using laser beam shaping strategy.
\newblock {\em Acta Materialia}, 184:284--305, 2020.

\bibitem{kirka2017strategy}
Michael~M Kirka, Yousub Lee, Duncan~A Greeley, Alfred Okello, Michael~J Goin, Michael~T Pearce, and Ryan~R Dehoff.
\newblock Strategy for texture management in metals additive manufacturing.
\newblock {\em Jom}, 69:523--531, 2017.

\bibitem{raplee2017thermographic}
J~Raplee, A~Plotkowski, Michael~M Kirka, R~Dinwiddie, A~Okello, Ryan~R Dehoff, and Sudarsanam~Suresh Babu.
\newblock Thermographic microstructure monitoring in electron beam additive manufacturing.
\newblock {\em Scientific reports}, 7(1):43554, 2017.

\bibitem{gould2021situ}
Benjamin Gould, Sarah Wolff, Niranjan Parab, Cang Zhao, Maria~Cinta Lorenzo-Martin, Kamel Fezzaa, Aaron Greco, and Tao Sun.
\newblock In situ analysis of laser powder bed fusion using simultaneous high-speed infrared and x-ray imaging.
\newblock {\em Jom}, 73:201--211, 2021.

\bibitem{wang2022situ}
Rongxuan Wang, David Garcia, Rakesh~R Kamath, Chaoran Dou, Xiaohan Ma, Bo~Shen, Hahn Choo, Kamel Fezzaa, Hang~Z Yu, and Zhenyu Kong.
\newblock In situ melt pool measurements for laser powder bed fusion using multi sensing and correlation analysis.
\newblock {\em Scientific reports}, 12(1):13716, 2022.

\bibitem{heigel2017effect}
Jarred~C Heigel and Brandon~M Lane.
\newblock The effect of powder on cooling rate and melt pool length measurements using in-situ thermographic techniques.
\newblock 2017.

\bibitem{heigel2018measurement}
Jarred~C Heigel and Brandon~M Lane.
\newblock Measurement of the melt pool length during single scan tracks in a commercial laser powder bed fusion process.
\newblock {\em Journal of Manufacturing Science and Engineering}, 140(5):051012, 2018.

\bibitem{cheng2019computational}
Bo~Cheng, Lukas Loeber, Hannes Willeck, Udo Hartel, and Charles Tuffile.
\newblock Computational investigation of melt pool process dynamics and pore formation in laser powder bed fusion.
\newblock {\em Journal of Materials Engineering and Performance}, 28:6565--6578, 2019.

\bibitem{tapia2014review}
Gustavo Tapia and Alaa Elwany.
\newblock A review on process monitoring and control in metal-based additive manufacturing.
\newblock {\em Journal of Manufacturing Science and Engineering}, 136(6):060801, 2014.

\bibitem{everton2016review}
Sarah~K Everton, Matthias Hirsch, Petros Stravroulakis, Richard~K Leach, and Adam~T Clare.
\newblock Review of in-situ process monitoring and in-situ metrology for metal additive manufacturing.
\newblock {\em Materials \& Design}, 95:431--445, 2016.

\bibitem{heigel2020situ}
Jarred~C Heigel, Brandon~M Lane, and Lyle~E Levine.
\newblock In situ measurements of melt-pool length and cooling rate during 3d builds of the metal am-bench artifacts.
\newblock {\em Integrating Materials and Manufacturing Innovation}, 9:31--53, 2020.

\bibitem{myers2023high}
Alexander~J Myers, Guadalupe Quirarte, Francis Ogoke, Brandon~M Lane, Syed~Zia Uddin, Amir~Barati Farimani, Jack~L Beuth, and Jonathan~A Malen.
\newblock High-resolution melt pool thermal imaging for metals additive manufacturing using the two-color method with a color camera.
\newblock {\em Additive Manufacturing}, 73:103663, 2023.

\bibitem{myers2023two}
Alexander~J Myers, Guadalupe Quirarte, Jack~L Beuth, and Jonathan~A Malen.
\newblock Two-color thermal imaging of the melt pool in powder-blown laser-directed energy deposition.
\newblock {\em Additive Manufacturing}, 78:103855, 2023.

\bibitem{moylan2014infrared}
Shawn Moylan, Eric Whitenton, Brandon Lane, and John Slotwinski.
\newblock Infrared thermography for laser-based powder bed fusion additive manufacturing processes.
\newblock In {\em AIP Conference Proceedings}, volume 1581, pages 1191--1196. American Institute of Physics, 2014.

\bibitem{hooper2018melt}
Paul~A Hooper.
\newblock Melt pool temperature and cooling rates in laser powder bed fusion.
\newblock {\em Additive Manufacturing}, 22:548--559, 2018.

\bibitem{zagade2021analytical}
P~Zagade, BP~Gautham, A~De, and T~DebRoy.
\newblock Analytical estimation of fusion zone dimensions and cooling rates in part scale laser powder bed fusion.
\newblock {\em Additive Manufacturing}, 46:102222, 2021.

\bibitem{heigel2020situ2}
Jarred~C Heigel, Brandon Lane, Lyle Levine, Thien Phan, and Justin Whiting.
\newblock In situ thermography of the metal bridge structures fabricated for the 2018 additive manufacturing benchmark test series (am-bench 2018).
\newblock {\em Journal of Research of the National Institute of Standards and Technology}, 125, 2020.

\bibitem{philips2020new}
NR~Philips, M~Carl, and NJ~Cunningham.
\newblock New opportunities in refractory alloys.
\newblock {\em Metallurgical and Materials Transactions A}, 51:3299--3310, 2020.

\bibitem{awasthi2022mechanical}
Prithvi~D Awasthi, Priyanka Agrawal, Ravi~Sankar Haridas, Rajiv~S Mishra, Michael~T Stawovy, Scott Ohm, and Aidin Imandoust.
\newblock Mechanical properties and microstructural characteristics of additively manufactured c103 niobium alloy.
\newblock {\em Materials Science and Engineering: A}, 831:142183, 2022.

\bibitem{colon2024parameter}
Brandon~J Col{\'o}n, Kurtis~I Watanabe, Toren~J Hobbs, Carly~J Romnes, Omar~R Mireles, Lawrence~E Murr, and Francisco Medina.
\newblock Parameter development and characterization of laser powder directed energy deposition of nb--alloy c103 for thin wall geometries.
\newblock {\em Journal of Materials Research and Technology}, 30:5028--5039, 2024.

\bibitem{mullin2024rapid}
Kaitlyn~M Mullin, Carolina Frey, James Lamb, Sophia~K Wu, McLean~P Echlin, and Tresa~M Pollock.
\newblock Rapid screening of single phase refractory alloys under laser melting conditions.
\newblock {\em Materials \& Design}, 238:112726, 2024.

\bibitem{mireles2020additive}
Omar Mireles, Omar Rodriguez, Youping Gao, and Noah Philips.
\newblock Additive manufacture of refractory alloy c103 for propulsion applications.
\newblock In {\em AIAA Propulsion and Energy 2020 Forum}, page 3500, 2020.

\bibitem{philips2024electron}
Noah Philips, Christopher Rock, Nicholas Cunningham, Josh Cooper, and Tim Horn.
\newblock Electron beam powder bed fusion of ati c103tm refractory alloy.
\newblock {\em Metallurgical and Materials Transactions A}, pages 1--13, 2024.

\bibitem{miklas2022additive}
Abigail Miklas.
\newblock {\em Additive Manufacturing of Refractory Alloys}.
\newblock PhD thesis, Colorado School of Mines, 2022.

\bibitem{satya2017niobium}
VV~Satya~Prasad, RG~Baligidad, and Amol~A Gokhale.
\newblock Niobium and other high temperature refractory metals for aerospace applications.
\newblock {\em Aerospace Materials and Material Technologies: Volume 1: Aerospace Materials}, pages 267--288, 2017.

\bibitem{cagran2009normal}
C~Cagran, H~Reschab, R~Tanzer, W~Sch{\"u}tzenh{\"o}fer, Andreas Graf, and Gernot Pottlacher.
\newblock Normal spectral emissivity of the industrially used alloys nicr20tial, inconel 718, x2crnimo18-14-3, and another austenitic steel at 684.5 nm.
\newblock {\em International journal of thermophysics}, 30:1300--1309, 2009.

\bibitem{shao2019grain}
Jiayun Shao, Gang Yu, Xiuli He, Shaoxia Li, Ru~Chen, and Yao Zhao.
\newblock Grain size evolution under different cooling rate in laser additive manufacturing of superalloy.
\newblock {\em Optics \& Laser Technology}, 119:105662, 2019.

\bibitem{farshidianfar2016effect}
Mohammad~H Farshidianfar, Amir Khajepour, and Adrian~P Gerlich.
\newblock Effect of real-time cooling rate on microstructure in laser additive manufacturing.
\newblock {\em Journal of Materials Processing Technology}, 231:468--478, 2016.

\bibitem{thampy2020subsurface}
Vivek Thampy, Anthony~Y Fong, Nicholas~P Calta, Jenny Wang, Aiden~A Martin, Philip~J Depond, Andrew~M Kiss, Gabe Guss, Qingfeng Xing, Ryan~T Ott, et~al.
\newblock Subsurface cooling rates and microstructural response during laser based metal additive manufacturing.
\newblock {\em Scientific reports}, 10(1):1981, 2020.

\bibitem{prasad2020towards}
Arvind Prasad, Lang Yuan, Peter Lee, Mitesh Patel, Dong Qiu, Mark Easton, and David StJohn.
\newblock Towards understanding grain nucleation under additive manufacturing solidification conditions.
\newblock {\em Acta Materialia}, 195:392--403, 2020.

\bibitem{van2007solutions}
Maarten Van~Elsen, Martine Baelmans, Peter Mercelis, and J-P Kruth.
\newblock Solutions for modelling moving heat sources in a semi-infinite medium and applications to laser material processing.
\newblock {\em International Journal of heat and mass transfer}, 50(23-24):4872--4882, 2007.

\bibitem{forslund2019analytical}
Robert Forslund, Anders Snis, and Stig Larsson.
\newblock Analytical solution for heat conduction due to a moving gaussian heat flux with piecewise constant parameters.
\newblock {\em Applied Mathematical Modelling}, 66:227--240, 2019.

\bibitem{mullin2024cracking}
Kaitlyn~M Mullin, Sebastian~A Kube, Sophia~K Wu, and Tresa~M Pollock.
\newblock Cracking and precipitation behavior of refractory bcc--b2 alloys under laser melting conditions.
\newblock {\em Metallurgical and Materials Transactions A}, pages 1--15, 2024.

\bibitem{fernandez2021crystallographic}
Patxi Fernandez-Zelaia, Christopher Ledford, Elizabeth~AI Ellis, Quinn Campbell, Andr{\'e}s~M{\'a}rquez Rossy, Donovan~N Leonard, and Michael~M Kirka.
\newblock Crystallographic texture evolution in electron beam melting additive manufacturing of pure molybdenum.
\newblock {\em Materials \& Design}, 207:109809, 2021.

\bibitem{thijs2013strong}
Lore Thijs, Maria Luz~Montero Sistiaga, Ruben Wauthle, Qingge Xie, Jean-Pierre Kruth, and Jan Van~Humbeeck.
\newblock Strong morphological and crystallographic texture and resulting yield strength anisotropy in selective laser melted tantalum.
\newblock {\em Acta Materialia}, 61(12):4657--4668, 2013.

\bibitem{gong2014beam}
Xibing Gong, James Lydon, Kenneth Cooper, and Kevin Chou.
\newblock Beam speed effects on ti--6al--4v microstructures in electron beam additive manufacturing.
\newblock {\em Journal of Materials Research}, 29(17):1951--1959, 2014.

\bibitem{chai2021effect}
Rongxia Chai, Yapu Zhang, Bin Zhong, and Chuanwei Zhang.
\newblock Effect of scan speed on grain and microstructural morphology for laser additive manufacturing of 304 stainless steel.
\newblock {\em Reviews on Advanced Materials Science}, 60(1):744--760, 2021.

\bibitem{kobryn2003microstructure}
Pamela~A Kobryn and SL~Semiatin.
\newblock Microstructure and texture evolution during solidification processing of ti--6al--4v.
\newblock {\em Journal of Materials Processing Technology}, 135(2-3):330--339, 2003.

\bibitem{liu2019insight}
Pengwei Liu, Zhuo Wang, Yaohong Xiao, Mark~F Horstemeyer, Xiangyang Cui, and Lei Chen.
\newblock Insight into the mechanisms of columnar to equiaxed grain transition during metallic additive manufacturing.
\newblock {\em Additive Manufacturing}, 26:22--29, 2019.

\end{thebibliography}




\newpage
\section*{Supporting Information}

\textbf{1. Single-track laser scanning and infrared measurement setup}

\begin{figure}[h!]
\centerline{\includegraphics[width=5in]
{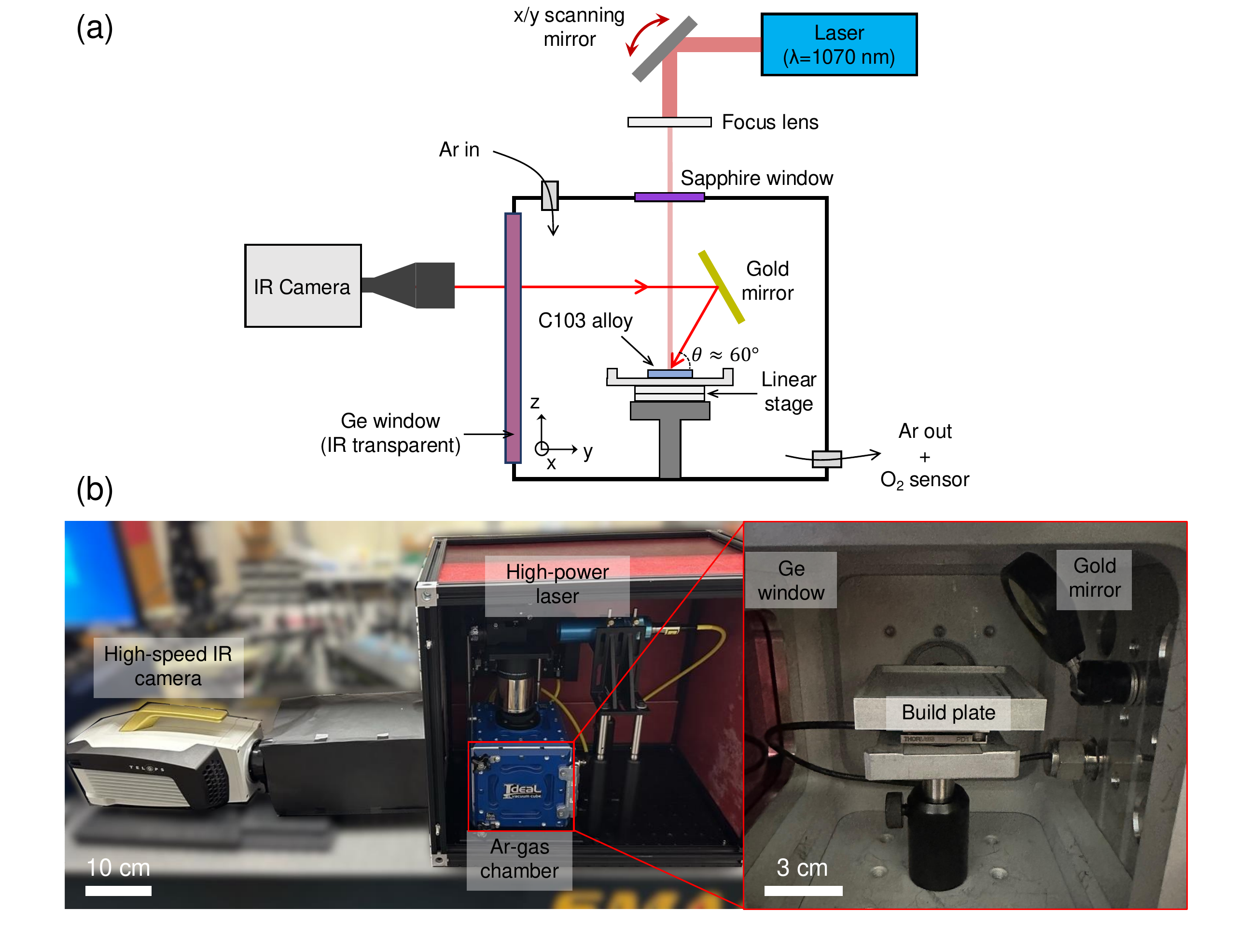}}
%\caption{A custom laser scanning setup interfaced with a high-speed IR camera. (a) a schematic of a custom laser processing chamber and optical components to interface with an infrared camera. (b) a photographic image of the custom chamber and infrared camera, with an inset detailing the mirror mount inside the side plate of the chamber. }
\textbf{Figure\,S1:} A custom laser scanning setup interfaced with a high-speed IR camera. (a) a schematic of a custom laser processing chamber and optical components to interface with an infrared camera. (b) a photographic image of the custom chamber and infrared camera, with an inset detailing the mirror mount inside the side plate of the chamber.

\label{setup}
\end{figure}

Figure\,S1 shows the schematic (Figure\,S1(a)) and photographic image (Figure\,S1(b)) of the custom Argon laser processing chamber interfaced with a high-speed IR camera. The custom laser processing chamber has been previously developed by Mullin et al. and the details can be found in reference \cite{mullin2024rapid}. Dynamic temperature maps during laser scanning were captured using a high-speed IR camera (Telops M3K) equipped with a 1X lens (Telops) with a frame rate of 15,136\,fps.  The IR camera can capture up to 100,000\,fps. However, at this extremely high frame rate, the window size is limited to 64\,px$\times$4\,px (1.92\,mm$\times$0.12\,mm). To capture the entire melt pool and track its movement, a moderate window size (3.84\,mm$\times$2.18\,mm) was selected, with a corresponding frame rate of 15,136 fps which is enough to capture the solidification dynamics. The IR camera is positioned adjacent to the laser scanning chamber. It views the sample through a germanium (Ge) window (transparent between 3-5 $\mu$m) and an IR-reflecting gold mirror, allowing the recording of thermal radiation emitted from the melt pool during laser scanning.  The received signal via the IR camera is in the form of radiant intensity, necessitating an accurate conversion of the intensity maps into temperature fields before further thermal analysis.

\clearpage

\textbf{2. Transmittance spectra of the Ge window}

\begin{figure}[h!]
\centerline{\includegraphics[width=5in]
{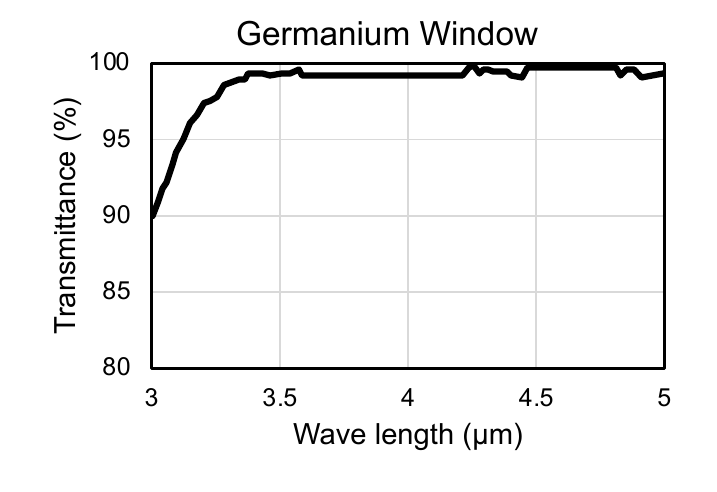}}
%\caption{Transmittance spectra of the Ge window. The high-speed IR camera (Telops M3K) has a spectral range of 3--5.5\,$\mu$m and the germanium (Ge) window is mostly transparent within this range. The non-unity transmittance of the Ge window especially near 3\,$\mu$m is accounted for in the calibration and is reflected by the effective emissivity. The transmission data was provided by the Ge window manufacturer (Sinoptix Optical).}
\textbf{Figure\,S2:} Transmittance spectra of the Ge window. The high-speed IR camera (Telops M3K) has a spectral range of 3--5.5\,$\mu$m and the germanium (Ge) window is mostly transparent within this range. The non-unity transmittance of the Ge window especially near 3\,$\mu$m is accounted for in the calibration and is reflected by the effective emissivity. The transmission data was provided by the Ge window manufacturer (Sinoptix Optical).

\label{Ge}
\end{figure}

\clearpage

\clearpage

\textbf{3. Temperature conversion}
\hfill \break

\begin{figure}[h!]
\centerline{\includegraphics[width=6in]
{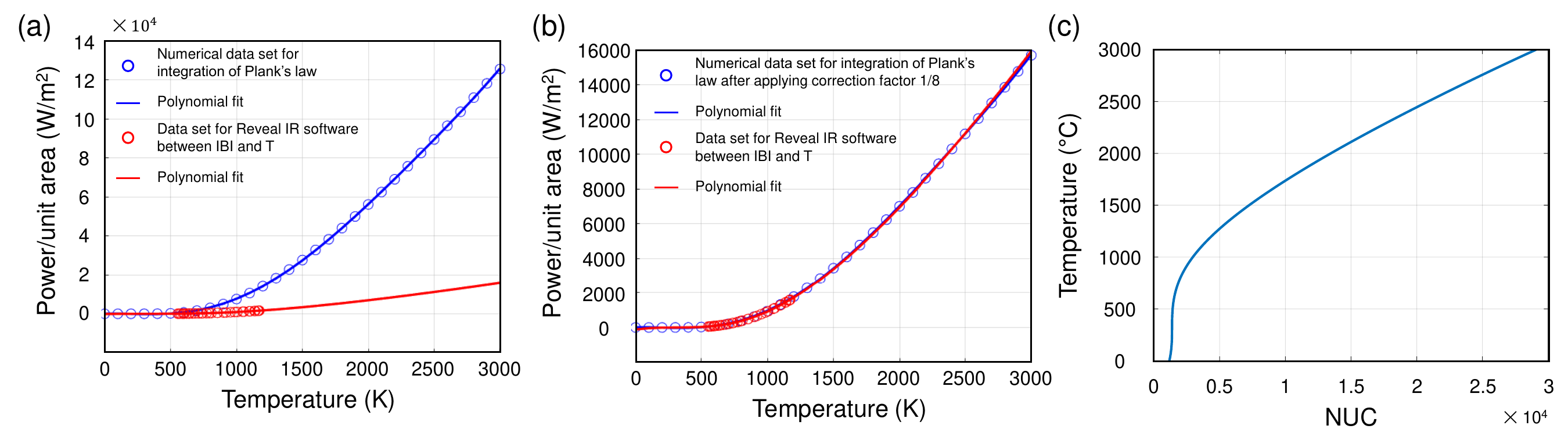}}
%\caption{Calibration curve from raw data to temperature}
\textbf{Figure\,S3:} Calibration curve from raw data to temperature

\label{SI_conversion}
\end{figure}

The theoretical irradiance incident on the camera detector's surface, referred to as the In-Band Irradiance (IBI) value (measured in W/m$^2$), is calculated from Plank's law as shown below.

\begin{equation}
L_\lambda(T) = \frac{2hc^2}{\lambda^5} \cdot \frac{1}{\exp\left(\frac{hc}{\lambda k_B T}\right) - 1}
\label{plank}
\end{equation} 

where $L_\lambda(T)$ is a spectral radiance at wavelength $\lambda$ , $h\!=\!$6.626\,$\times \!10^{-34}$ J$\cdot$s is the Planck's constant, $c\!=\!3\times 10^8$\,m/s is the speed of light in a vacuum, $\lambda$ is a wavelength of the emitted radiation ($\mu$m), $k_B\!=\,1.381\,\times 10^{-23}$\,J/K is the Boltzmann constant, and $T$ is the absolute temperature of the blackbody (K). By integrating Planck's law, which governs the spectral radiation of a black body, over the 3-5.5 $\mu$m wavelength range, the theoretical spectral radiation is derived. The corresponding values, along with temperature, are plotted as blue data points in Figure\,S3(a).

To calculate the in-band irradiance received by the camera from the theoretical spectral radiation, several factors must be considered, including the camera's tilt angle, view factor, and the camera's solid angle. The IR camera manufacturer, Telops, provides a calibration to account for these parameters, converting temperature to In-Band Irradiance (IBI). The relationship between temperature and IBI is shown in Figure\,S3(a) as red data points.

There is a noticeable gap between the theoretical spectral radiation obtained by integrating over the 3-5.5$\mu$m wavelength range (blue data points) and the in-band irradiance values provided by the IR camera manufacturer (red data points). To address this discrepancy, a correction factor of 1/8 is applied, which is specific to our camera and the camera's position (as explained in Section S1). After applying this correction factor to the theoretical spectral radiation (blue data), it aligns with the IBI (red data) values provided by the manufacturer, as shown in Figure\,S3(b). The IBI and temperature relationship provides an approximate description of how the IR raw signal varies with the temperature of a black body.

Factors such as the sample's non-unity in-band emissivity within 3-5.5\,$\mu$m, the sensor's non-ideal spectral absorbance, and the optics' transmittance in the IR camera necessitate additional corrections to convert the theoretical IBI to the actual detected signal (NUC). The camera software (Reveal IR) includes a built-in correction to perform this conversion, taking into account camera acquisition parameters like exposure time and frame rate. Details on how the NUC and IBI are related under our experimental condition, are described in the main text equation Equation\,1. Figure\,S3(c) shows the final result of the temperature vs NUC curve.

\clearpage

\textbf{4. Scanning electron microscope image}
\hfill \break

\begin{figure}[h!]
\centerline{\includegraphics[width=5.5in]
{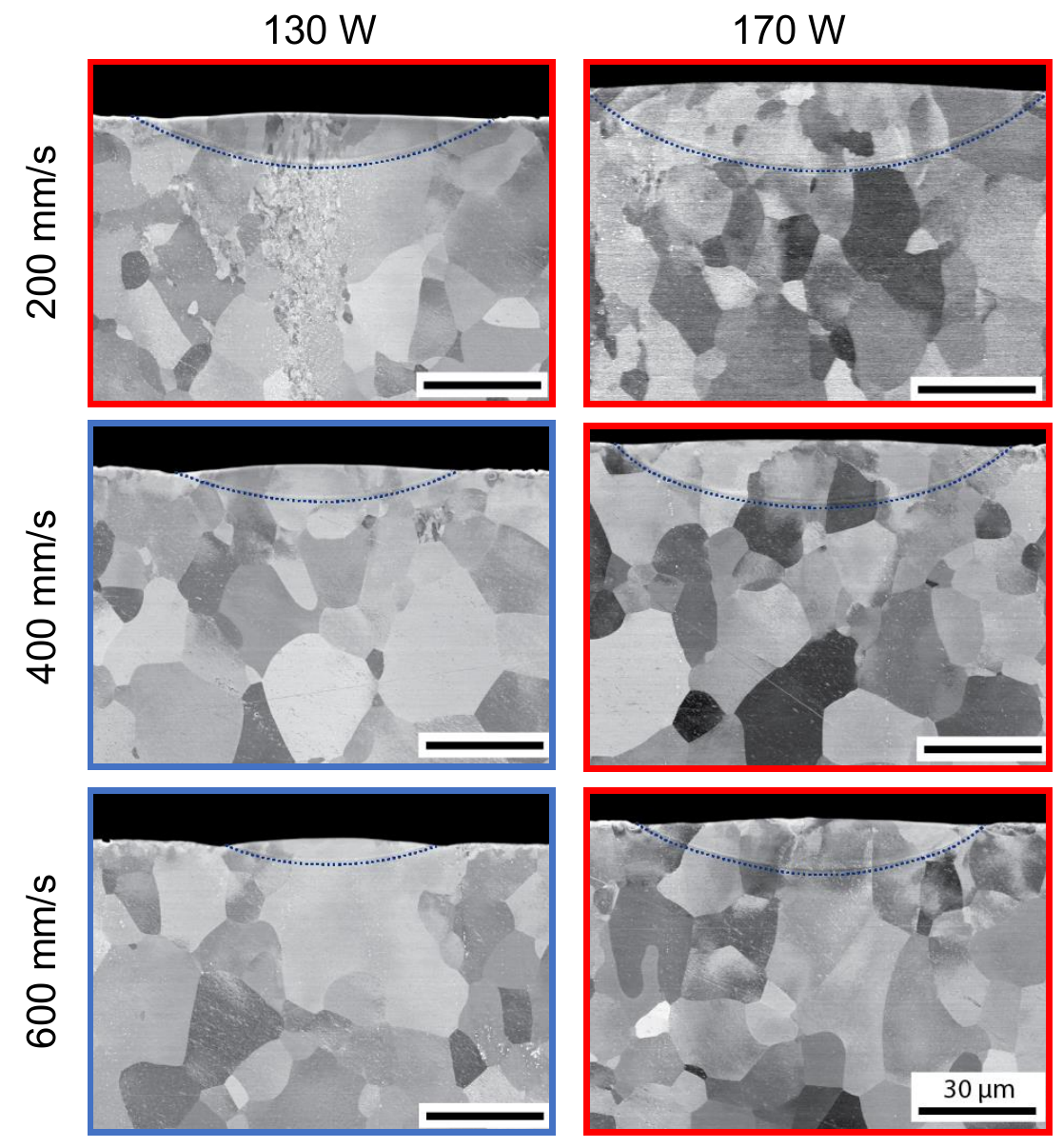}}
%\caption{SEM images of the cross-section of the melt pool. The wider melt pool width shows nucleated grains within the melt pool (red boxes), in contrast, only epitaxial growth from the melt pool boundary was observed for melt pools with narrower widths (blue boxes).}
\textbf{Figure\,S4:} SEM images of the cross-section of the melt pool. The wider melt pool width shows nucleated grains within the melt pool (red boxes), in contrast, only epitaxial growth from the melt pool boundary was observed for melt pools with narrower widths (blue boxes).

\label{SEM}
\end{figure}

\clearpage

\textbf{5. Melt pool shape}
\hfill \break

\begin{figure}[h!]
\centerline{\includegraphics[width=6in]
{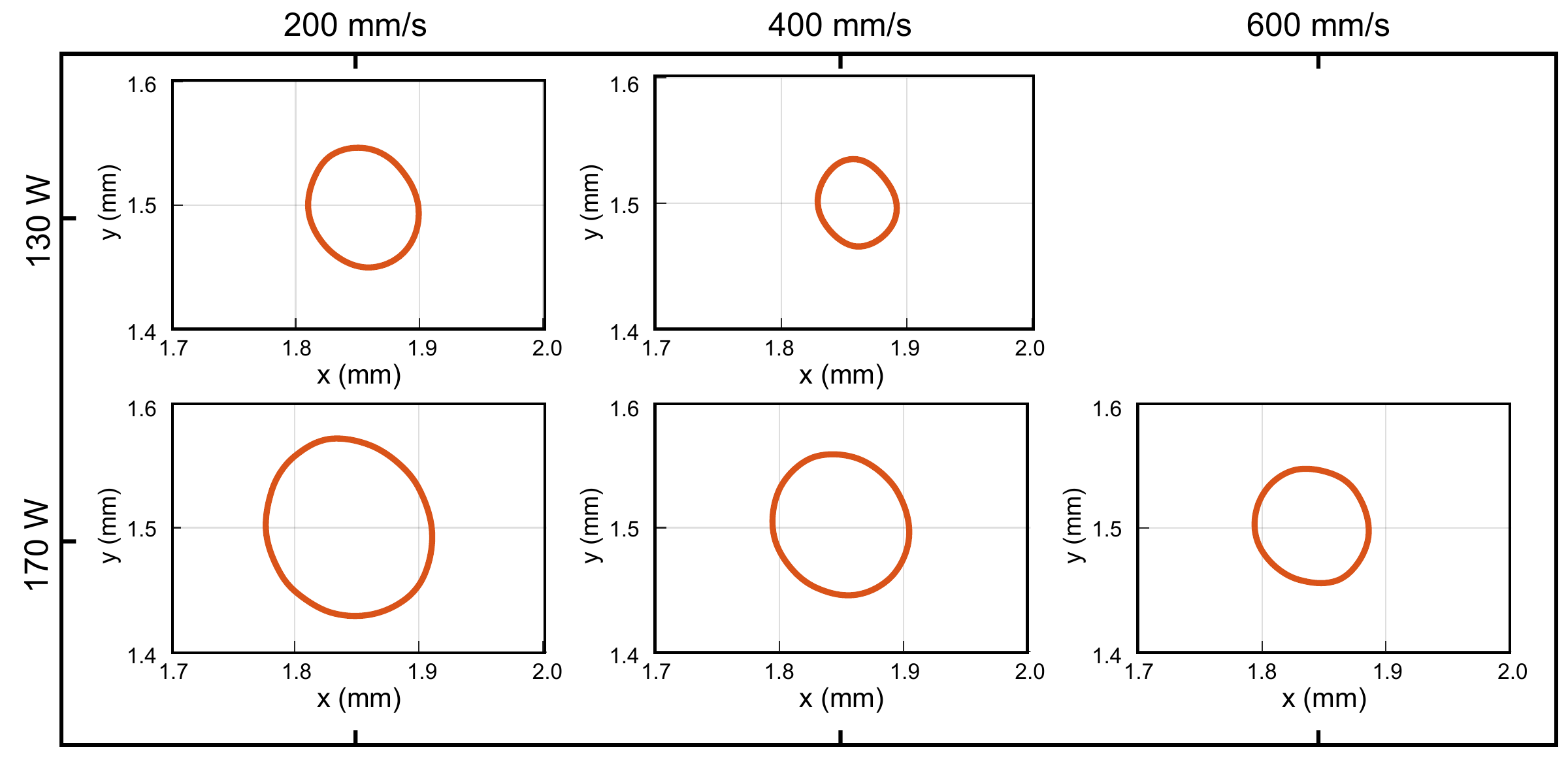}}
%\caption{The solid-liquid boundary shape of each experiment is determined using the method explained in Figure\,4 in the main text. The higher power (170\,W) results in a larger melt pool size compared to the lower power laser (130\,W) at the same scan speed.  Also, the slower speed (200\,mm/s) yields a larger melt pool size than the higher speed (600\,mm/s) at the same laser power. Due to a hardware issue, the IR recording of the 130 W 600 mm/s was not saved correctly.}

\textbf{Figure\,S5:} The solid-liquid boundary shape of each experiment is determined using the method explained in Figure\,4 in the main text. The higher power (170\,W) results in a larger melt pool size compared to the lower power laser (130\,W) at the same scan speed.  Also, the slower speed (200\,mm/s) yields a larger melt pool size than the higher speed (600\,mm/s) at the same laser power. Due to a hardware issue, the IR recording of the 130 W 600 mm/s was not saved correctly.

\label{SI_S-L boundary}
\end{figure}

\clearpage

\textbf{6. Isothermal line velocity}
\hfill \break

\begin{figure}[h!]
\centerline{\includegraphics[width=3in]
{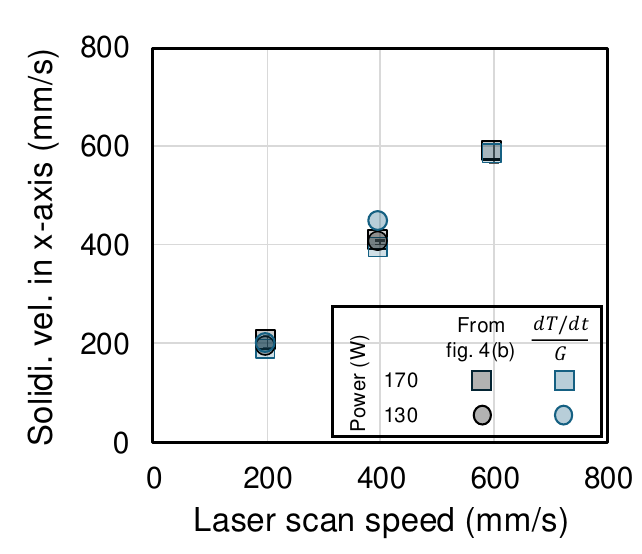}}
%\caption{Comparison of the melt pool moving velocity measured from Fig. 4(b) with the solidification velocity of the trailing edge of the melt pool calculated using Equation\,5 in the main text(black data).}
\textbf{Figure\,S6:} Comparison of the melt pool moving velocity measured from Fig. 4(b) with the solidification velocity of the trailing edge of the melt pool calculated using Equation\,5 in the main text(black data).

\label{Center Sol Vel}
\end{figure}

The cooling rate ($dT/ds$) and the temperature gradient ($G$) of a point on the melt pool boundary can be used to calculate the solidification velocity. As explained in the main text, typically the total temperature gradient includes a contribution from the $z$-direction. Since in this paper, $G$ is the projected temperature gradient on the $x$-$y$ plane, the solidification velocity calculated is the projected solidification velocity on the $x$-$y$ plane. For the trailing edge on the melt pool boundary,   $R_x$ (blue data) is the solidification velocity calculated from Equation\,5 in the main manuscript, which represents the isothermal line velocity. The results align well with the isothermal line velocity obtained by directly tracking the movement of the melt pool boundary, as indicated in Figure\,4(b) of the main text (black data point).

\clearpage

\textbf{7. Temperature gradient calculation}
\hfill \break

Solidification occurs when the temperature drops to the solidus temperature during cooling, assuming minimal subcooling, which needs further verification. The magnitude of the temperature gradient on the $x$--$y$ plane is

 \begin{equation}
G = \sqrt{\left(\frac{dT}{dx}\right)^2+\left(\frac{dT}{dy}\right)^2},
\label{gradient}
\end{equation}

The total temperature gradient in 3D should include a $z$-direction component. However, since the IR camera only captures temperature maps on the $x$-$y$ plane with no $z$-direction information, the temperature gradient we are considering is only the projected gradient on the $x$-$y$ plane. For the temperature gradient along the center laser scan path,  only the $x$-direction temperature gradient $G_x$ is needed due to symmetry in the $y$-direction,  so Equation\,\ref{gradient} reduced to

\begin{equation}
G_x = \frac{dT}{dx} \approx \frac{\Delta T_{x}}{\Delta x},
\label{G_x}
\end{equation}

where $\Delta x$ is the distance between adjacent points in the
smoothed temperature map in the $x$-direction (3\,$\mu$m after smoothing the temperature map), and $\Delta T_{x}$ is estimated using a linear approximation of the temperature centered difference method which is

\begin{equation}
\Delta T_{x} = \frac{\mid T_x-T_{x-1}\mid +\mid T_x-T_{x+1}\mid }{2},
\label{delta_x}
\end{equation}

In Equation\,\ref{delta_x}, $x$ is the corresponding point location in the $x$-direction of the image, and $T_{x+1}$ and $T_{x-1}$ are the temperatures of adjacent pixels. The discretized temperatures (every 3\,$\mu$m) were obtained by interpolating the original temperature data points, which have a coarser mesh, using a spline smooth function in MATLAB.

For the temperature gradient map on the melt pool boundary, the numerical form of equation \ref{gradient} is:

 \begin{equation}
G \approx \sqrt{\left(\frac{\Delta T_{x}}{\Delta x}\right)^2+\left(\frac{\Delta T_{y}}{\Delta y}\right)^2},
\label{G_spatial}
\end{equation}

where $\Delta x$ and $\Delta y$ are the distance between adjacent points in the smoothed temperature map in the $x$- and $y$-directions (3\,$\mu$m and 3.4\,$\mu$m, respectively, after smoothing the temperature map using spline interpolation in MATLAB), and $\Delta T_{x}$ and $\Delta T_{y}$ are estimated using a linear approximation of the temperature centered difference method. Similar to Equation\,\ref{delta_x}, $\Delta T_{y}$  is calculated as follows:

\begin{equation}
\Delta T_{y} = \frac{\mid T_y-T_{y-1}\mid +\mid T_y-T_{y+1}\mid }{2},
\label{delta_y}
\end{equation}

In Equation\,\ref{delta_y}, $y$ is the corresponding pixel location in the $y$-direction of the image, and $T_{y+1}$ and $T_{y-1}$ are the temperatures of adjacent pixels. Using these linear approximations, the gradient ($G$) is calculated and discussed in the main manuscript.

\clearpage

\textbf{8. $y$-direction solidification velocity}
\hfill \break

\begin{figure}[h!]
\centerline{\includegraphics[width=6in]
{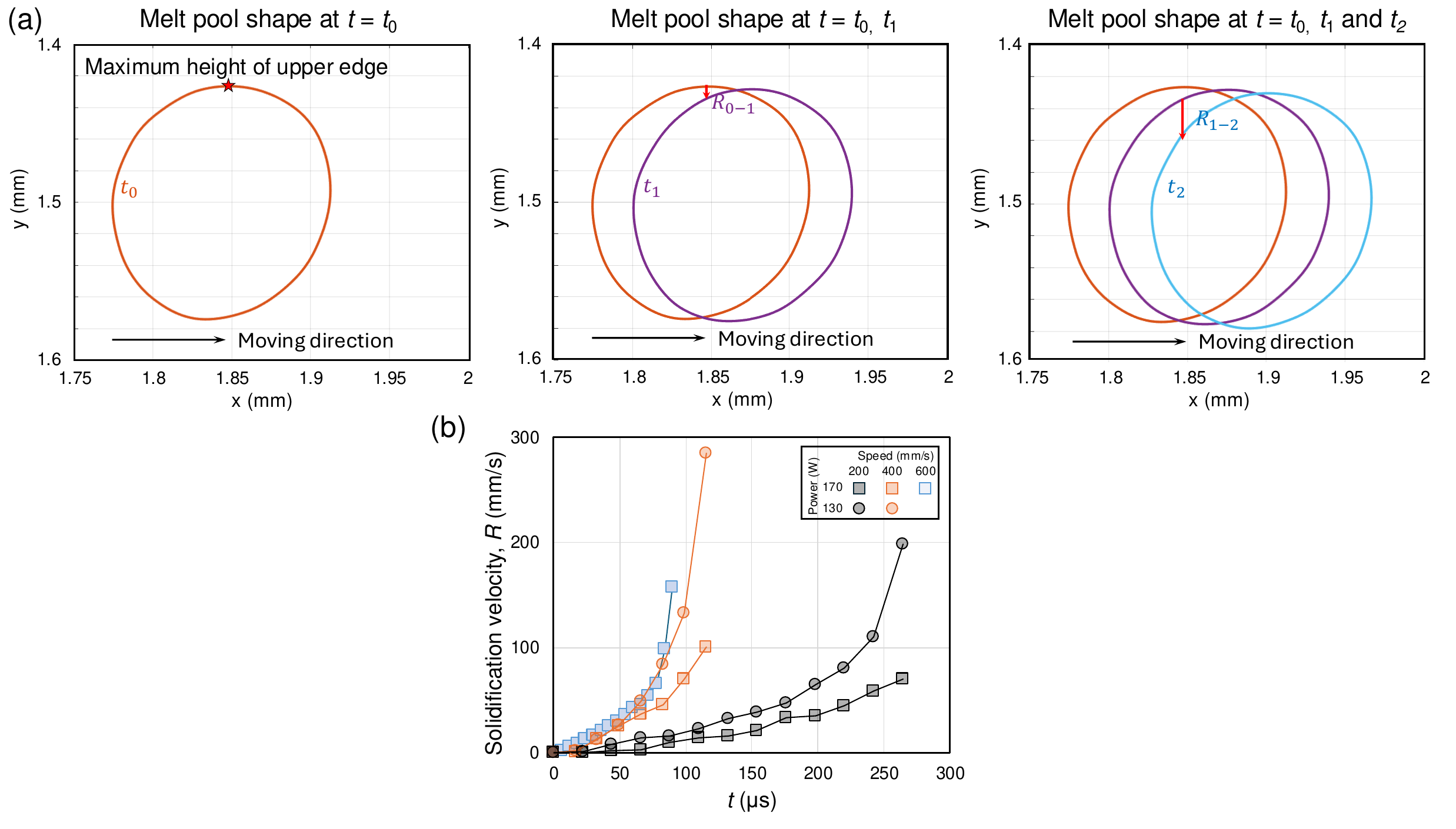}}
%\caption{Solidification velocity based on the melt pool shape.  (a) illustrates the measurement of the solidification velocity ($R$) from the outermost edge of the melt pool (indicated by the red star) to the center. (b) The solidification velocity ($R$)  is measured over time. }

\textbf{Figure\,S7:} Solidification velocity based on the melt pool shape.  (a) illustrates the measurement of the solidification velocity ($R$) from the outermost edge of the melt pool (indicated by the red star) to the center. (b) The solidification velocity ($R$)  is measured over time.

\label{SI_Solid volocity}
\end{figure}

The solidification velocity at the edges of the melt pool (along the $y$-direction) moving inward can also be calculated by tracking the melt pool boundary from the side edge to the center, perpendicular to the laser scanning direction. As the melt pool moves in the laser scanning direction in each time frame, the solidification velocity, $R$, from the outermost edge of the melt pool to its center is determined. Due to the rapid movement of the melt pool, the frame rate of our experiment does not provide sufficient solidification data measurements. To address this issue, we interpolated two (for 200\,mm/s scan speed), three (for 400\,mm/s scan speed), and ten (for 600\,mm/s scan speed) imaginary melt pool boundaries between adjacent time intervals of approximately  66\,$\mu$s (with 15,136 fps for our experiment). This interpolation allows for more precise solidification rate measurements with time steps of approximately 6-20\,$\mu$s in each frame. The solidification velocity was calculated as follows: At $t=t_\textbf{0}$, using the highest top position of the melt pool as a reference, the velocity from this maximum height at the upper edge to the melt pool center during the interval  $t_\textbf{0}$ and $t_\textbf{1}$ is measured as $R_\textbf{0-1}$. Similarly, the solidification velocity between $t_\textbf{1}$ and $t_\textbf{2}$, is calculated as $R_\textbf{1-2}$.

The $y$-direction solidification velocity as a function of time is presented in Figure\,S7(b). It can be seen that the solidification velocity accelerates as the edge moves towards the center. The time it takes for the edges to meet at the center (the last data point in Figure\,S7(b)) reduces as the laser scanning speed increases. The slower laser scan speed (black data) exhibits a slower solidification velocity compared to the faster scan speed (orange and blue data). In addition, higher laser power (square-shaped data) also results in a slower solidification rate compared to the lower laser power (circle-shaped data). This observation aligns with our expectation since slower scan speeds and higher laser power retain more heat, thereby prolonging the solidification time of the melt pool. Rapid increases in solidification velocity are observed at the tail end of solidification, particularly for melt tracks with lower powers.

% \end{thebibliography}
\end{document}